\documentclass[12pt]{article}
\usepackage{amsfonts}
\usepackage{epic}
\begin{document}
\pagestyle{empty}

\newcommand{\si}{\sigma}
\newcommand{\pr}{\prime}
\newcommand{\br}[1]{|#1\rangle}
\newcommand{\vs}[1]{\rule[- #1 mm]{0mm}{#1 mm}}
\newcommand{\hs}[1]{\hspace{#1 mm}}
\newcommand{\sm}[2]{\frac{\mbox{\footnotesize #1}\vs{-2}}
                   {\vs{-2}\mbox{\footnotesize #2}}}

\newcommand{\slq}{U_q sl_2}
\newcommand{\hr}{\hat{R}}
\newcommand{\D}{\Delta}
\newcommand{\ot}{\otimes}
\newcommand{\nn}{\nonumber}
\newcommand{\hot}{\hat{\otimes}}
\newcommand{\insi}{\stackrel{\displaystyle <}{\sim}}
\newcommand{\susi}{\stackrel{\displaystyle >}{\sim}}
\newcommand{\op}{\oplus}
\newcommand{\lgl}{\langle}
\def\CC{{\mathbb C}}
\def\NN{{\mathbb N}}
\def\QQ{{\mathbb Q}}
\def\RR{{\mathbb R}}
\def\ZZ{{\mathbb Z}}
\def\cA{{\cal A}}          \def\cB{{\cal B}}          \def\cC{{\cal C}}
\def\cD{{\cal D}}          \def\cE{{\cal E}}          \def\cF{{\cal F}}
\def\cG{{\cal G}}          \def\cH{{\cal H}}          \def\cI{{\cal I}}
\def\cJ{{\cal J}}          \def\cK{{\cal K}}          \def\cL{{\cal L}} 
\def\cM{{\cal M}}          \def\cN{{\cal N}}          \def\cO{{\cal O}}
\def\cP{{\cal P}}          \def\cQ{{\cal Q}}          \def\cR{{\cal R}} 
\def\cS{{\cal S}}          \def\cT{{\cal T}}          \def\cU{{\cal U}}
\def\cV{{\cal V}}          \def\cW{{\cal W}}          \def\cX{{\cal X}}
\def\cY{{\cal Y}}          \def\cZ{{\cal Z}}
\def\vR{\check{R}}
\def\qmbox#1{\qquad\mbox{#1}\qquad}
\def\Id{1\hspace{-1mm}{\sf I}}
 
\begin{center} 
{\Large \textsf{
Integrable Chain Model with Additional Staggered 
\\[4mm]
Model Parameter}} 

\vfill

{\bf D.~Arnaudon}\footnote{e-mail:{\sl arnaudon@lapp.in2p3.fr}},
{\bf R.~Poghossian}\footnote{e-mail:{\sl poghos@moon.yerphi.am,
Permanent address: Yerevan Physics Institute, Armenia}},
{\bf A.~Sedrakyan}\footnote{e-mail:{\sl sedrak@lx2.yerphi.am,
Permanent address: Yerevan Physics Institute, Armenia}},
{\bf P.~Sorba}\footnote{e-mail:{\sl sorba@lapp.in2p3.fr}}\\

\vfill
\emph{Laboratoire d'Annecy-le-Vieux de Physique Th{\'e}orique LAPTH}
\\
\emph{CNRS, UMR 5108, associ{\'e}e {\`a} l'Universit{\'e} de Savoie}
\\
\emph{LAPP, BP 110, F-74941 Annecy-le-Vieux Cedex, France}
\\


\vfill
{\bf Abstract}
\end{center}
The generalization of the Yang-Baxter equations (YBE) in the presence
of $\ZZ_2$ grading along both chain and time directions is presented. 
The $XXZ$ model with staggered disposition along a chain of both, the
anisotropy $\pm \Delta$, as well as shifts of the spectral 
parameters are considered and the corresponding integrable model is 
constructed. The Hamiltonian of the model is computed in fermionic
and spin formulations. It involves three neighbour site interactions
and therefore can be considered as a zig-zag ladder model. The Algebraic
Bethe Ansatz technique is applied and the eigenstates, along with
eigenvalues of the transfer matrix of the model are found.
The model has a free fermionic limit at $\Delta=0$ and the integrable
boundary terms are found in this case.

This construction is quite general and can be applied to other 
known integrable models.

\vfill
\rightline{LAPTH-781/00}
\rightline{hep-th/0002123}
\rightline{February 2000}

\newpage
\pagestyle{plain}
\setcounter{page}{1}

\section{Introduction}
\indent
The construction of integrable chain models with inhomogeneous
model parameters is very important in many aspects. Besides its own
interest in mathematical physics, its possible applications in the 
problems of condensed matter physics is hard to miss-estimate. Even more,
we believe that in order to proceed further in the non-critical
string theory and cross the so called $c=1$ \cite{KPZ} barrier,
it is necessary to consider strongly correlated fields on a line,
instead of the free field degrees of freedom on the world
sheet of the string (as in $d\leq1$ case) .

The crucial problem for strings in $d \geq 1$ is the problem
of the tachyon, which means a non-trivial structure of the ground state.
Therefore, the investigation of strongly correlated chain systems, 
which can be considered on a fluctuating line, is actual.

In this sense, in order to be able to take into account non-trivial
ground state correctly, it is straightforward to consider an
integrable $2d$ model on a fluctuating surface. One attractive
possibility is the consideration of inhomogeneously
shifted spectral parameters in $R$-matrices along a chain. It is 
tempting to consider first the case of staggered inhomogeneity.

An interesting model of hopping fermions in the Manhattan lattice,
where the hopping parameters are staggered in both time and chain
directions, appeared in connection with three dimensional Ising
Model  \cite{KS, S}. In a particular case of hopping 
parameters, this produces a $N \times N$ numerical transfer matrix
\cite{CC} in connection with edge excitations  
in the Hall effect.

Since the pioneering work of H.~Bethe \cite{B} and the essential input 
made by C.~Yang \cite{Y}, the theory of $2d$-integrable models
is well developed presently and has its formulation in technique, 
called Algebraic Bethe Ansatz (ABA) (or Quantum Inverse
Scattering Method(QISM)) \cite{Kor, Bax}.

The basic element of this method is the matrix $R_{i j}(\lambda, u)$
($\lambda$ is the model parameter and $u$ is the spectral parameter),
the product of which along a chain defines the monodromy matrix 
$T(\lambda,u)$ as follows
\begin{equation}
\label{M}
T(\lambda,u)=\prod_{j=1}^N R_{0 j}(\lambda, u).
\end{equation}
The trace of $T(\lambda,u)$ over states along the chain defines the transfer
matrix $\tau(\lambda,u)$, the $L$'s power of which ($L$ is the
lattice size in time direction) is the partition function of the model.

When any two transfer matrices of the model with different spectral
parameters commute then we have an infinite number of conservation
currents, which causes the integrability. The sufficient condition
of commutativity of the transfer matrices is the Yang-Baxter
equation (YBE) for the $R$-matrix, any solution of which defines an
integrable model.

Usually the integrable models under consideration are homogeneous 
along a chain, namely, the spectral parameters $u$ in the product 
(\ref{M}) are equal. There were attempts to construct 
inhomogeneous models, for example we can mention the successful 
attempt in the article \cite{Vega} of the model with 
alternating spin-${1 \over 2}$, spin-$1$ chain, but the main problem
here is to obtain a local Hamiltonian (local conservation
laws). It is obvious that, with the shift of the spectral parameter
$u$ in the product of $R_{0j}(\lambda,u)$ matrices at the site $j$
by arbitrary values $z_j$, we still fulfill the YBE (and hence
the commutativity of transfer matrices) however the Hamiltonian
of the model, and the other conservation quantities, will 
be local only if the shifts have periodic structure.
Due to periodicity which goes far from neighbour sites,
the local Hamiltonian will contain interactions between sites
within the period and therefore can be represented as multi-leg
ladder model. This type of ladder models differs from ones,
constructed in \cite{Wa} by extension of the symmetry algebra,
in \cite{Ko} by construction first of the ground state,
in \cite{Muu} by use of higher order conservation laws and
others \cite{Al, For}.

The aim of the present article is to construct an integrable 
model with commuting transfer matrices for different spectral 
parameters, where the $R$-matrices in the product (\ref{M})
are staggered by their construction as well as by
shift of the spectral parameter $u$ and where the Hamiltonian
(and the other conservation laws) is  local. The shift of the spectral
parameter produces an additional model parameter $\theta$.
Similar model, where only the shift of spectral parameter
was considered (no change of the structure of the $R$-matrix),
was constructed earlier in \cite{ZV, FR}.

We have generalized the YBE for that purpose in order to ensure
the commutativity of the transfer matrices and demonstrate that 
they do have a solution in a case when the basic $R$-matrix is the one
of the integrable $XXZ$-model.

In Section 2 the generalized YBE is formulated and the 
solution is found for the basic $XXZ$-model.

In Section 3 we  calculate the Hamiltonian for the $XXZ$ model 
in both, fermionic and spin formulations, and
demonstrate that it is local. 
Because of the invariance of the model under double space
translations, the Hamiltonian contains interaction between three
neighbouring points  
(instead of nearest-neighbour links in the ordinary $XXZ$).
It can hence be represented as a zig-zag ladder Hamiltonian,
where we have two terms, corresponding to $SU(1,1)$ Heisenberg models 
on  each of two chains, 
and the interaction between chains terms of topological form, which 
are written on the triangles. 
In the free fermionic limit ($\lambda=\pi/2$) we have obtained 
two independent sets of fermions, hopping
only at the odd or even sites. 

This method, as one in \cite{ZV,FR},
provides a way of construction of the integrable ladder
models.

Algebraic Bethe Ansatz ($ABA$)
technique is applied to present model in Section 4 in order to 
find the eigenstates and eigenvalues of the transfer matrix.

In Section 5 the open chain problem is considered and
integrable reflection $\cal K$-matrices are found for the
staggered $XX$-model. We also write the corresponding
boundary terms in the Hamiltonian.

We think that our approach for the construction of 
integrable models with staggered model parameters
is quite general and can be applied to other integrable 
homogeneous models.


\section{The definition of the YBE for the staggered model
and its solution for the $XXZ$ chain}
\indent

The key of integrability of the models is the YBE, which implies 
some restrictions on the $R_{ij}(\lambda,u)$-matrix, basic constituent
of the monodromy  matrix (\ref{M}). 
The YBE ensures a local sufficient
condition for the commutativity of the transfer-matrices 
$\tau (u)= tr T(u)$ at different values 
of the
spectral parameter $u$, which corresponds to the rapidity of
pseudoparticles of the model
\begin{equation}
\label{1.11}
[\tau \left( u\right),\tau \left( v\right)]=0.
\end{equation}

We use in the future the fermionization
technique (alternative to Jordan-Wigner transformation), developed
in \cite{AHS, AKS, GM} (see also \cite{USW}) and work with the 
$R$-operators (rather than matrices), expressed in Fermi-fields
$c_i$, $c_i^+$, ($i =1,....N$ is the chain site).

By definition $R_{aj}$ acts as a intertwining operator on the 
space of direct product of
the so called auxiliary $V_{a}\left( v\right) $ and quantum $V_{j}\left(
u\right) $ spaces 
\begin{equation}
\label{R1}
R_{aj}\left( u,v\right):\quad V_{a}(u)\otimes V_{j}(v)\rightarrow
V_{j}(v)\otimes V_{a}(u)  
\end{equation}
and can be represented graphically as
\begin{center}
  {\small {\ \setlength{\unitlength}{5mm} 
      \begin{picture}(10,15)
        \put(3.9,4.5){$V_j(v)$}
        \put(0,8){\vector(1,0){7}}
        \put(0.5,8.5){$V_a(u)$}
        \put(3.5,10.6){\vector(0,1){0.7}}
        \put(3,10.8){\line(-1,-1){0.2}}
        \put(3,10.8){\line(1,-1){0.2}}
        \multiput(3.5,5.4)(0,1){6}{\line(0,-1){0.7}}
        \put(-5,7.85){$R_{aj}(u,v)=$} 
        \put(0,3){\shortstack{Figure 1: $R_{aj}$ matrix}}
      \end{picture}} }
\end{center}
\label{fig:1}
\addtocounter{figure}{1}

The spaces $V_a(u)$ and $V_j(v)$ with spectral parameters $u$ and $v$ are
irreducible representations of the affine quantum group $U_q\widehat{g},$
which is the symmetry group of the integrable model under consideration.
Provided that the states $\mid a\rangle $ $\in V_a$ and $\mid j\rangle \in
V_j$ form a basis for the spaces $V_a$ and $V_j$, following \cite{AHS}
we can represent the action of the operator $R_{aj}$ as  
\begin{equation}
\label{R2}
R_{aj}\mid j_1\rangle \otimes \mid a_1\rangle = 
\left(R_{aj}\right)^{a_2 j_2}_{a_1 j_1}
\mid a_2\rangle \otimes \mid j_2\rangle , 
\end{equation}
where the summation is over the repeating indices $a_2$ and
$j_2$ (but not over $a_1$ and $j_1$).

By introducing the Hubbard operators 
\begin{equation}
\label{X1}
X_{a_2}^{a_1}=\mid a_2\rangle \langle {a_1}\mid,  \qquad
X_{j_2}^{j_1}=\mid j_2\rangle \langle {j_1}\mid  
\end{equation}
in the graded spaces $V_a$ and $V_j$ correspondingly, one can 
rewrite (\ref{R2}) as 
\begin{eqnarray}
\label{R3}
R_{aj}=R_{aj}\mid {j_1}\rangle \mid {a_1}\rangle \langle {a_1}\mid
\langle {j_1}\mid 
&=&
\left(R_{aj}\right)^{a_2 j_2}_{a_1 j_1}
\mid a_2\rangle
\mid j_2\rangle \langle {a_1}\mid \langle {j_1}\mid  \nonumber\\
&=&
\left(
-1\right) ^{p({a_1})p(j_2)}
\left(R_{aj}\right)^{a_2j_2}_{{a_1}{j_1}}X_{a_2}^{a_1} X_{j_2}^{j_1}  
\end{eqnarray}
where the sign factor takes into account the possible grading of 
the states $\mid
a_i\rangle $ and $\mid j_i\rangle ,$ $p(a_i)$ and $p(j_i)$ denote the 
corresponding
parities and the summation over the repeating indices.

In terms of operators $R_{ij}$ the matrix valued YBE can be written
in the following operator form
\begin{equation}
\label{RRR}
R_{a b}(u,v)R_{aj}(u,w)R_{bj}(v,w)=R_{bj}(v,w)R_{aj}(u,w)R_{a b}(u,v).  
\end{equation}

We use for $V_0$ and $V_j$ the Fock space 
of the Fermi fields $\footnote{if the dimensions of the spases $V_0$ and
$V_j$ are more than two, then one can use Fock spaces of more fermions,
see \cite{AKS}}$
$c_i$, $c_i^+$ with basis vectors $| 0 \rangle_i$ and 
$| 1 \rangle_i$, for which
\begin{equation}
\label{X2}
(X_i)_a^{a^{\pr}}=\left(\begin{array}{ll}
1-n_i & c_i^+ \nn\\
c_i & n_i
\end{array}\right),
\end{equation}
we have a fermionization of the model, which is equivalent
to Jordan-Wigner transformation.

In case of $XXZ$ model, the fermionic expression of the operator 
$R_{ij}(u)$ is easy to obtain by use of formulas (\ref{R3}) and (\ref{X2}),
and the standard expression for the $R_{ba}^{b^{\pr} a^{\pr}}(u)$. 
As a result one obtains 
\begin{eqnarray}
\label{R4}
R_{jk}(u)&=&a(u)\left[-n_j n_k +(1-n_j)(1-n_k)\right]+
b(u)\left[n_j(1- n_k) +(1-n_j) n_k\right]\nn\\
&+&c(u)\left[c_j^+ c_k + c_k^+ c_j\right].
\end{eqnarray}

Let us now consider $\ZZ_2$ graded quantum $V_{j,\rho}(v)$ and 
auxiliary $V_{a,\si}(u)$ spaces, $\rho, \si =0,1$. In this case 
we will have $4\times 4$ $R$-matrices, which act on the direct product
of the spaces $V_{a,\si}(u)$ and $ V_{j,\rho}(v)$, $(\si,\rho =0,1)$,
mapping them on the intertwined direct product of 
$V_{a,\bar{\si}}(u)$ and $ V_{j,\bar{\rho}(v)}$ with the complementary
$\bar{\si}=(1-\si)$, $\bar{\rho}=(1-\rho)$ indices
\begin{equation}
\label{R5}
R_{aj,\si \rho}\left( u,v\right):\quad V_{a,\si}(u)\otimes 
V_{j,\rho}(v)\rightarrow V_{j,\bar{\rho}}(v)\otimes V_{a,\bar{\si}}(u).  
\end{equation}

It is convenient to introduce two transmutation operations $\iota_1$
and $\iota_2$ with the property $\iota_1^2=\iota_2^2=id$ 
for the quantum and auxiliary spaces
correspondingly, and to mark the operators $R_{aj,\si\rho}$ as 
follows
\begin{eqnarray}
\label{R51}
R_{aj,00}&\equiv& R_{aj},\qquad R_{aj,01}\equiv R_{aj}^{\iota_1},\nn\\
R_{aj,10}&\equiv& R_{aj}^{\iota_2},\qquad R_{aj,11}\equiv R_{aj}^{\iota_1 
\iota_2}.
\end{eqnarray}

The introduction of the $\ZZ_2$ grading in quantum space means, 
that we have now two monodromy matrices $T_{\rho}, \rho=0,1$,
which act on the space $V_{\rho}=\prod_{j=1}^N V_{j,\rho}$
by mapping it on $V_{\bar{\rho}}=\prod_{j=1}^N V_{j,\bar{\rho}}$
\begin{equation}
\label{T2}
T_\rho \qquad : V_\rho \rightarrow V_{\bar{\rho}}, \qquad \qquad 
\rho=0,1.
\end{equation}

It is clear now, that the monodromy matrix of the model, which should 
define the partition function, is the product of two monodromy matrices
\begin{equation}
\label{T3}
T(\lambda,u) = T_0(\lambda,u) T_1(\lambda,u).
\end{equation}

Now, because of the grading in the auxiliary space, we would like to
construct the monodromy matrices $T_{0,1}$ as a staggered product
of the $R_{aj}$ and $\bar{R}_{aj}^{\iota_2}$ matrices.
Let us define
\begin{eqnarray}
\label{T4}
T_1(\lambda,u)=\prod_{j=1}^N R_{a,2j-1}(\lambda,u)
\bar{R}_{a,2j}^{\iota_2}(\lambda,u)\nn\\
T_0(\lambda,u)=\prod_{j=1}^N \bar{R}_{a,2j-1}^{\iota_1}(\lambda,u)
R_{a,2j}^{\iota_1 \iota_2}(\lambda,u),
\end{eqnarray}
where the notation $\bar{R}$ in general means the
different parameterization of the $R$-matrix via models
$\lambda$ and $u$ parameters and can be considered as an operation
over $R$ with property $\bar{\bar{R}}= R$.

Graphically the formulas (\ref{T4}) can be expressed as in 
figure \ref{fig:2}.
\def\comVa#1{{$V_{a,#1}$}}
\def\comVj#1{{$V_{j,#1}$}}
\def\comTa{{$T_1(\lambda,u)$}}
\def\comTb{{$T_0(\lambda,u)$}}
\begin{figure}[h]
  \begin{center}
    \leavevmode
\setlength{\unitlength}{0.00033333in}
\begingroup\makeatletter\ifx\SetFigFont\undefined
\def\x#1#2#3#4#5#6#7\relax{\def\x{#1#2#3#4#5#6}}%
\expandafter\x\fmtname xxxxxx\relax \def\y{splain}%
\ifx\x\y   
\gdef\SetFigFont#1#2#3{%
  \ifnum #1<17\tiny\else \ifnum #1<20\small\else
  \ifnum #1<24\normalsize\else \ifnum #1<29\large\else
  \ifnum #1<34\Large\else \ifnum #1<41\LARGE\else
     \huge\fi\fi\fi\fi\fi\fi
  \csname #3\endcsname}%
\else
\gdef\SetFigFont#1#2#3{\begingroup
  \count@#1\relax \ifnum 25<\count@\count@25\fi
  \def\x{\endgroup\@setsize\SetFigFont{#2pt}}%
  \expandafter\x
    \csname \romannumeral\the\count@ pt\expandafter\endcsname
    \csname @\romannumeral\the\count@ pt\endcsname
  \csname #3\endcsname}%
\fi
\fi\endgroup
{\renewcommand{\dashlinestretch}{30}
\begin{picture}(16212,11895)(0,-10)
\drawline(3000,3036)(5400,3036)
\drawline(5280.000,3006.000)(5400.000,3036.000)(5280.000,3066.000)
\drawline(5400,3036)(7800,3036)
\drawline(7680.000,3006.000)(7800.000,3036.000)(7680.000,3066.000)
\drawline(7800,3036)(10200,3036)
\drawline(10080.000,3006.000)(10200.000,3036.000)(10080.000,3066.000)
\drawline(10200,3036)(12600,3036)
\drawline(12480.000,3006.000)(12600.000,3036.000)(12480.000,3066.000)
\drawline(12600,3036)(15000,3036)
\drawline(14880.000,3006.000)(15000.000,3036.000)(14880.000,3066.000)
\drawline(15000,3036)(16200,3036)
\drawline(6000,561)(6300,1761)
\drawline(6300.000,1637.307)(6300.000,1761.000)(6241.791,1651.859)
\drawline(6300,1761)(6900,4161)
\drawline(6900.000,4037.307)(6900.000,4161.000)(6841.791,4051.859)
\drawline(6900,4161)(7200,5361)
\drawline(10800,561)(11100,1761)
\drawline(11100.000,1637.307)(11100.000,1761.000)(11041.791,1651.859)
\drawline(11100,1761)(11700,4161)
\drawline(11700.000,4037.307)(11700.000,4161.000)(11641.791,4051.859)
\drawline(11700,4161)(12000,5361)
\drawline(4800,561)(4500,1761)
\drawline(4558.209,1651.859)(4500.000,1761.000)(4500.000,1637.307)
\drawline(4500,1761)(3900,4161)
\drawline(3958.209,4051.859)(3900.000,4161.000)(3900.000,4037.307)
\drawline(3900,4161)(3600,5361)
\drawline(9600,561)(9300,1761)
\drawline(9358.209,1651.859)(9300.000,1761.000)(9300.000,1637.307)
\drawline(9300,1761)(8700,4161)
\drawline(8758.209,4051.859)(8700.000,4161.000)(8700.000,4037.307)
\drawline(8700,4161)(8400,5361)
\drawline(14400,561)(14100,1761)
\drawline(14158.209,1651.859)(14100.000,1761.000)(14100.000,1637.307)
\drawline(14100,1761)(13500,4161)
\drawline(13558.209,4051.859)(13500.000,4161.000)(13500.000,4037.307)
\drawline(13500,4161)(13200,5361)
\drawline(3600,6636)(3900,7836)
\drawline(3900.000,7712.307)(3900.000,7836.000)(3841.791,7726.859)
\drawline(3900,7836)(4500,10236)
\drawline(4500.000,10112.307)(4500.000,10236.000)(4441.791,10126.859)
\drawline(4500,10236)(4800,11436)
\drawline(8400,6636)(8700,7836)
\drawline(8700.000,7712.307)(8700.000,7836.000)(8641.791,7726.859)
\drawline(8700,7836)(9300,10236)
\drawline(9300.000,10112.307)(9300.000,10236.000)(9241.791,10126.859)
\drawline(9300,10236)(9600,11436)
\drawline(13200,6636)(13500,7836)
\drawline(13500.000,7712.307)(13500.000,7836.000)(13441.791,7726.859)
\drawline(13500,7836)(14100,10236)
\drawline(14100.000,10112.307)(14100.000,10236.000)(14041.791,10126.859)
\drawline(14100,10236)(14400,11436)
\drawline(1800,9036)(3000,9036)
\drawline(2880.000,9006.000)(3000.000,9036.000)(2880.000,9066.000)
\drawline(3000,9036)(5400,9036)
\drawline(5280.000,9006.000)(5400.000,9036.000)(5280.000,9066.000)
\drawline(5400,9036)(7800,9036)
\drawline(7680.000,9006.000)(7800.000,9036.000)(7680.000,9066.000)
\drawline(7800,9036)(10200,9036)
\drawline(10080.000,9006.000)(10200.000,9036.000)(10080.000,9066.000)
\drawline(10200,9036)(12600,9036)
\drawline(12480.000,9006.000)(12600.000,9036.000)(12480.000,9066.000)
\drawline(12600,9036)(15000,9036)
\drawline(14880.000,9006.000)(15000.000,9036.000)(14880.000,9066.000)
\drawline(15000,9036)(16200,9036)
\drawline(1800,3036)(3000,3036)
\drawline(2880.000,3006.000)(3000.000,3036.000)(2880.000,3066.000)
\drawline(7200,6636)(6900,7836)
\drawline(6958.209,7726.859)(6900.000,7836.000)(6900.000,7712.307)
\put(3000,11436){\makebox(0,0)[lb]{\smash{{{\SetFigFont{12}{14.4}{rm}\comVj0}}}}}
\drawline(6900,7836)(6300,10236)
\drawline(6358.209,10126.859)(6300.000,10236.000)(6300.000,10112.307)
\drawline(6300,10236)(6000,11436)
\drawline(12000,6636)(11700,7836)
\drawline(11758.209,7726.859)(11700.000,7836.000)(11700.000,7712.307)
\drawline(11700,7836)(11100,10236)
\drawline(11158.209,10126.859)(11100.000,10236.000)(11100.000,10112.307)
\drawline(11100,10236)(10800,11436)
\put(2100,8436){\makebox(0,0)[lb]{\smash{{{\SetFigFont{12}{14.4}{rm}\comVa0}}}}}
\put(4800,8436){\makebox(0,0)[lb]{\smash{{{\SetFigFont{12}{14.4}{rm}\comVa1}}}}}
\put(7200,8436){\makebox(0,0)[lb]{\smash{{{\SetFigFont{12}{14.4}{rm}\comVa0}}}}}
\put(9900,8436){\makebox(0,0)[lb]{\smash{{{\SetFigFont{12}{14.4}{rm}\comVa1}}}}}
\put(12225,8436){\makebox(0,0)[lb]{\smash{{{\SetFigFont{12}{14.4}{rm}\comVa0}}}}}
\put(14400,8436){\makebox(0,0)[lb]{\smash{{{\SetFigFont{12}{14.4}{rm}\comVa1}}}}}
\put(2100,2436){\makebox(0,0)[lb]{\smash{{{\SetFigFont{12}{14.4}{rm}\comVa0}}}}}
\put(4800,2436){\makebox(0,0)[lb]{\smash{{{\SetFigFont{12}{14.4}{rm}\comVa1}}}}}
\put(7200,2436){\makebox(0,0)[lb]{\smash{{{\SetFigFont{12}{14.4}{rm}\comVa0}}}}}
\put(9900,2436){\makebox(0,0)[lb]{\smash{{{\SetFigFont{12}{14.4}{rm}\comVa1}}}}}
\put(12225,2436){\makebox(0,0)[lb]{\smash{{{\SetFigFont{12}{14.4}{rm}\comVa0}}}}}
\put(14400,2436){\makebox(0,0)[lb]{\smash{{{\SetFigFont{12}{14.4}{rm}\comVa1}}}}}
\put(6000,11736){\makebox(0,0)[lb]{\smash{{{\SetFigFont{12}{14.4}{rm}2}}}}}
\put(9600,11736){\makebox(0,0)[lb]{\smash{{{\SetFigFont{12}{14.4}{rm}3}}}}}
\put(10800,11736){\makebox(0,0)[lb]{\smash{{{\SetFigFont{12}{14.4}{rm}4}}}}}
\put(14400,11736){\makebox(0,0)[lb]{\smash{{{\SetFigFont{12}{14.4}{rm}5}}}}}
\put(7200,6036){\makebox(0,0)[lb]{\smash{{{\SetFigFont{12}{14.4}{rm}2}}}}}
\put(8400,6036){\makebox(0,0)[lb]{\smash{{{\SetFigFont{12}{14.4}{rm}3}}}}}
\put(12075,6036){\makebox(0,0)[lb]{\smash{{{\SetFigFont{12}{14.4}{rm}4}}}}}
\put(13200,6036){\makebox(0,0)[lb]{\smash{{{\SetFigFont{12}{14.4}{rm}5}}}}}
\put(4500,11736){\makebox(0,0)[lb]{\smash{{{\SetFigFont{12}{14.4}{rm}j=1}}}}}
\put(3300,6036){\makebox(0,0)[lb]{\smash{{{\SetFigFont{12}{14.4}{rm}j=1}}}}}
\put(4500,36){\makebox(0,0)[lb]{\smash{{{\SetFigFont{12}{14.4}{rm}j=1}}}}}
\put(6000,36){\makebox(0,0)[lb]{\smash{{{\SetFigFont{12}{14.4}{rm}2}}}}}
\put(9600,36){\makebox(0,0)[lb]{\smash{{{\SetFigFont{12}{14.4}{rm}3}}}}}
\put(10800,36){\makebox(0,0)[lb]{\smash{{{\SetFigFont{12}{14.4}{rm}4}}}}}
\put(14400,36){\makebox(0,0)[lb]{\smash{{{\SetFigFont{12}{14.4}{rm}5}}}}}
\put(0,9036){\makebox(0,0)[lb]{\smash{{{\SetFigFont{12}{14.4}{rm}\comTa}}}}}
\put(0,3036){\makebox(0,0)[lb]{\smash{{{\SetFigFont{12}{14.4}{rm}\comTb}}}}}
\put(3000,636){\makebox(0,0)[lb]{\smash{{{\SetFigFont{12}{14.4}{rm}\comVj0}}}}}
\put(1800,5436)
{\makebox(0,0)[lb]{\smash{{{\SetFigFont{12}{14.4}{rm}\comVj1}}}}}
\put(1800,6636)
{\makebox(0,0)[lb]{\smash{{{\SetFigFont{12}{14.4}{rm}\comVj1}}}}}
\end{picture}
}
    
  \end{center}
  \label{fig:2}
  \caption{Monodromy matrices $T_0$ and $T_1$}
\end{figure}
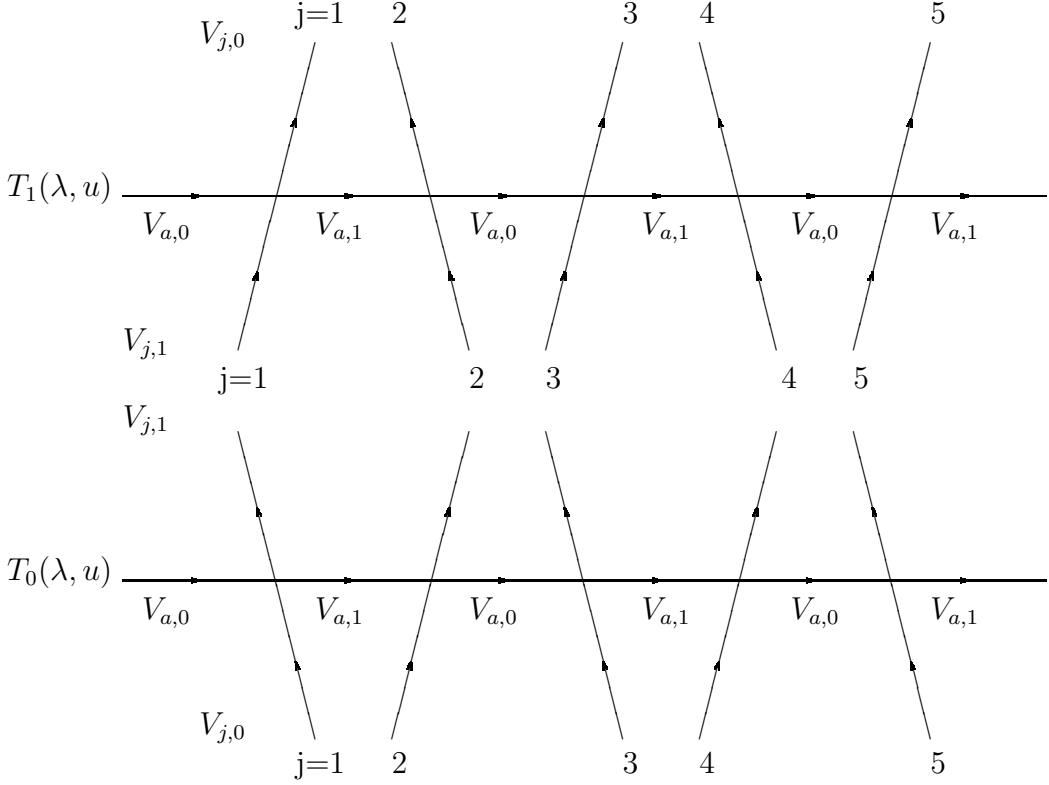

In order to have a integrable model with commuting transfer matrices
(\ref{T3}) for different spectral parameters
\begin{equation}
\label{T5}
\left[ trT(\lambda,u), \; trT(\lambda,v)\right]=0
\end{equation}
it is enough to have the following relations for the $\tau_{\si}(u)=
trT_{\si}(\lambda,u), \qquad (\si=0,1)$
\begin{equation}
\label{T51}
\tau_{\si}(\lambda,u)\tau_{1-\si}(\lambda,v)=
\bar{\tau}_{\si}(\lambda,v)\bar{\tau}_{1-\si}(\lambda,u),\qquad \si=0,1
\end{equation}

It is not hard to see, that in order to ensure the commutativity
condition (\ref{T5}) the $R$- and $\bar{R}$-matrices in the expression 
(\ref{T4}) should fulfill the following two Yang-Baxter
equations
\begin{eqnarray}
\label{R6}
R_{ab}(u,v) \bar{R}_{a,j}^{\iota_1}(\lambda,u) R_{b,j}(\lambda,v)=
R_{b,j}^{\iota_1}(\lambda,v) \bar{R}_{a,j}(\lambda,u) \tilde{R}_{ab}(u,v)
\end{eqnarray}
and
\begin{eqnarray}
\label{R7}
\tilde{R}_{ab}(u,v) R_{a,j+1}^{\iota_1 \iota_2}(\lambda,u) 
\bar{R}_{b,j+1}^{\iota_2}(\lambda,v)=
\bar{R}_{b,j+1}^{\iota_1 \iota_2}(\lambda,v) R_{a,j+1}^{\iota_2}(\lambda,u) R_{ab}(u,v).
\end{eqnarray}
By inserting the expression (\ref{R4}) for the $R$-operator
into (\ref{R6}) one can obtain 20 equations on parameters $a$, $b$, $c$
and $\bar{a}$, $\bar{b}$, $\bar{c}$, which after defining the 
transposition operation $\iota_1$ and $\sim$ as
\begin{eqnarray}
\label{abcT1}
a^{\iota_1}(u) &=& a(u), \qquad b^{\iota_1}(u) = -b(u), \qquad c^{\iota_1}=c(u)\nn\\
\tilde{a}(u,v) &=& a(u,v), \qquad \tilde{b}(u,v) = -b(u,v), 
\qquad \tilde{c}(u,v) = c(u,v),
\end{eqnarray}
are reducing to following three equations
\begin{eqnarray}
\label{abc1}
a(u,v) \bar{c}(u) a(v) = c(u,v) \bar{a}(u) c(v) + 
b(u,v) \bar{c}(u) b(v),\nn\\
b(u,v) \bar{a}(u) c(v) + c(u,v) \bar{c}(u) b(v) = 
a(u,v) \bar{b}(u) c(v),\nn\\
c(u,v) \bar{b}(u) a(v) = b(u,v) \bar{c}(u) c(v) + 
c(u,v) \bar{a}(u) b(v). 
\end{eqnarray}

It is easy to recognize in equations (\ref{abc1}) the ordinary
YBE for the $XXZ$ model, where we suppose that the parametrization
of the variables $\bar{a}(u)$, $\bar{b}(u)$, $\bar{c}(u) $
differs from $a(u)$, $b(u)$, $c(u)$. We are going to introduce a new
model parameter here for them.

The second set of YBEs (\ref{R7}) gives
\begin{eqnarray}
\label{abc2}
a(u,v) c^{\iota_2}(u) \bar{a}^{\iota_2}(v) = c(u,v) {a}^{\iota_2}(u) \bar{c}^{\iota_2}(v)  
-b(u,v) {c}^{\iota_2}(u) \bar{b}^{\iota_2}(v),\nn\\
-b(u,v) {a}^{\iota_2}(u) \bar{c}^{\iota_2}(v) + c(u,v) {c}^{\iota_2}(u) \bar{b}^{\iota_2}(v) 
= a(u,v) {b}^{\iota_2}(u) \bar{c}^{\iota_2}(v),\nn\\
c(u,v) {b}^{\iota_2}(u) \bar{a}^{\iota_2}(v) = -b(u,v) {c}^{\iota_2}(u) \bar{c}^{\iota_2}(v) 
+c(u,v) {a}^{\iota_2}(u) \bar{b}^{\iota_2}(v). 
\end{eqnarray} 
which differs from the first set (\ref{abc1}) by the transposition
$\iota_2$ of the variables, by displaced bars from the  
$\bar{a}(u)$, $\bar{b}(u)$, $\bar{c}(u) $ on to
$a(v)$, $b(v)$, $c(v)$ and change of sign $b(u,v)\rightarrow -b(u,v)$.
This last change of the sign comes from the fact that, in the first set 
of YBEs (\ref{R6}), the variables $\tilde{b}(u,v)$ stationed in the right
side of equations, whilst in the second set (\ref{R7}) - in
the left side.

The solution of the equations (\ref{abc1}) is the same as for ordinary
$XXZ$, but we will make a different shifts of the spectral parameters
for the $\bar{a}(u)$, $\bar{b}(u)$, $\bar{c}(u)$ and $a(u)$, $b(u)$, $c(u)$.

Namely, if we take
\begin{eqnarray}
\label{Sol1}
a(u)&=& \sin(\lambda +u),\qquad \bar{a}(u) = a(\bar{u})=
\sin(\lambda+ \theta -u),\nn\\
b(u)&=& \sin(u), \qquad \qquad
\bar{b}(u) = b(\bar{u})=\sin(\theta -u),\nn\\
c(u)&=& \sin(\lambda), \qquad \qquad \bar{c}(u) = c(\bar{u})=
\sin(\lambda),
\end{eqnarray}
then the solution of (\ref{abc1}) for $a(u,v)$, $b(u,v)$, $c(u,v)$
is
\begin{eqnarray}
\label{Sol2}
a(u,v)&=& a(\bar{u}-v)=\sin(\lambda+\bar{\eta} -u-v),\nn\\
b(u,v)&=& b(\bar{u}-v)=\sin(\bar{\eta}-u-v),\nn\\
c(u,v)&=& c(\bar{u}-v)=\sin(\lambda),
\end{eqnarray}
where $\bar{u}=\theta -u$ and $\theta$ is the new  
parameter of the model.

It is easy to see after simple calculations, that the second set of 
YBEs (\ref{abc2}) is compatible with the first set (\ref{abc1})
by producing the same solution (\ref{Sol2}), if we define
the $\iota_2$ transposition as follows
\begin{eqnarray}
\label{abcT2}
a^{\iota_2}(u)&=&\sin(\lambda -u),\qquad b^{\iota_2}(u)=b(u),
\qquad c^{\iota_2}=c(u),\nn\\
\bar{a}^{\iota_2}(u)&=&\sin(\lambda-\bar{u}),\qquad 
\bar{b}^{\iota_2}(u)=b(\bar{u}), \qquad \bar{c}^{\iota_2}=c(\bar{u}).
\end{eqnarray}

Therefore, the formula (\ref{Sol1}) together with
the definitions (\ref{abcT1}) and (\ref{abcT2}) defines a solution
of the double set of YBEs (\ref{R6}) and (\ref{R7}). The new parameter
$\theta$ appeared in the model in addition
to $\lambda$ (or $\Delta$) in the ordinary $XXZ$-model.

As it is well known, the variable
\begin{equation}
\label{D}
\Delta={a^2(u)+b^2(u)-c^2(u) \over 2 a(u) b(u)}=\cos\lambda=const
\end{equation}
characterize the anisotropy of the product of the spins
$\vec{\si}_j$ and $\vec{\si}_{j+1} $ in $z$-direction
in the Heisenberg model.

It is important to mention here, that the transpositions $\iota_1$
and $\iota_2$ changes the sign of $\Delta$, $\Delta_{\iota_1}=
\Delta_{\iota_2}=- \Delta$. This means, that in a definition
(\ref{T4}) of the monodromy matrices $T_{0,1}(\lambda,u)$ we
have a product of $R$-matrices with alternating anisotropy
parameters $\pm \Delta$. This fact essentially differs our construction
from the one in the articles \cite{ZV,FR}, where the simple shift
of the spectral parameter is considered.

\section{The Hamiltonian of the model. Ladder form}
\indent

The prove of existence and the calculation of the local
Hamiltonian in our model is easy to carry out technically
in a so called check-formalism, where instead of usual
$R$-matrix, we consider the matrix $\check{R}= P R$,
($P$ being the permutation map 
$P= x \otimes y \rightarrow y \otimes x$). In matrix elements
\begin{equation}
\label{Ch}
\check{R}_{a b}^{a^{\pr} b^{\pr}}= R_{a b}^{b^{\pr} a^{\pr}}.
\end{equation}

Instead of the operator $R_{aj}$ in  (\ref{R3}) let us define
\begin{equation}
\label{R31}
\check{R}_{ij}=
(\check{R}_{ij})_{a b}^{a^{\pr}b^{\pr}} \mid a^{\prime }\rangle
\mid b^{\prime }\rangle \langle b\mid \langle a\mid =\left(
-1\right) ^{p(a^{\pr})(p(b^{\prime })+p(b))}
(\check{R}_{ij})_{a b}^{a^{\pr}b^{\pr}}
X_{a^{\prime}}^a X_{b^{\prime }}^b ,  
\end{equation}
which after the fermionization (\ref{X2}) will have the form
\begin{eqnarray}
\label{R41}
\check{R}_{jk}(u)&=&a(u)\left[n_j n_k +(1-n_j)(1-n_k)\right]+
c(u)\left[n_j(1- n_k) +(1-n_j) n_k\right]\nn\\
&+&b(u)\left[c_j^+ c_k + c_k^+ c_j\right].
\end{eqnarray}
and where the parameters $a(u)$, $b(u)$, $c(u)$ are defined by formulas 
(\ref{Sol1}).

The monodromy matrix (\ref{T4}) in this formalism has the 
following expression
\begin{eqnarray}
\label{T41}
T_1(\lambda,u)=\prod_{j=1}^N \check{R}_{2j,2j+1}(\lambda,u)
\check{\bar{R}}_{2j+1,2j+2}^{\iota_2}(\lambda,u),\nn\\
T_0(\lambda,u)=\prod_{j=1}^N \check{\bar{R}}_{2j+1,2j+2}^{\iota_1}(\lambda,u)
\check{R}_{2j+2,2j+3}^{\iota_1 \iota_2}(\lambda,u).
\end{eqnarray}

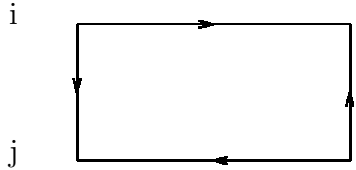
\begin{figure}[htbp]
  \begin{center}
    \leavevmode
\setlength{\unitlength}{0.0006in}
\begingroup\makeatletter\ifx\SetFigFont\undefined
\def\x#1#2#3#4#5#6#7\relax{\def\x{#1#2#3#4#5#6}}%
\expandafter\x\fmtname xxxxxx\relax \def\y{splain}%
\ifx\x\y   
\gdef\SetFigFont#1#2#3{%
  \ifnum #1<17\tiny\else \ifnum #1<20\small\else
  \ifnum #1<24\normalsize\else \ifnum #1<29\large\else
  \ifnum #1<34\Large\else \ifnum #1<41\LARGE\else
     \huge\fi\fi\fi\fi\fi\fi
  \csname #3\endcsname}%
\else
\gdef\SetFigFont#1#2#3{\begingroup
  \count@#1\relax \ifnum 25<\count@\count@25\fi
  \def\x{\endgroup\@setsize\SetFigFont{#2pt}}%
  \expandafter\x
    \csname \romannumeral\the\count@ pt\expandafter\endcsname
    \csname @\romannumeral\the\count@ pt\endcsname
  \csname #3\endcsname}%
\fi
\fi\endgroup
{\renewcommand{\dashlinestretch}{30}
\begin{picture}(3012,1359)(0,-10)
\drawline(600,1236)(1800,1236)
\drawline(1680.000,1206.000)(1800.000,1236.000)(1680.000,1266.000)
\drawline(3000,36)(3000,636)
\drawline(3030.000,516.000)(3000.000,636.000)(2970.000,516.000)
\drawline(600,1236)(3000,1236)(3000,36)
        (600,36)(600,1236)
\drawline(3000,36)(1800,36)
\drawline(1920.000,66.000)(1800.000,36.000)(1920.000,6.000)
\put(0,36){\makebox(0,0)[lb]{\smash{{{\SetFigFont{12}{14.4}{rm}j}}}}}
\drawline(600,1236)(600,636)
\drawline(570.000,756.000)(600.000,636.000)(630.000,756.000)
\put(0,1236){\makebox(0,0)[lb]{\smash{{{\SetFigFont{12}{14.4}{rm}i}}}}}
\end{picture}
}
  \end{center}
  \caption{$\check{R}_{ij}$}
  \label{fig:3}
\end{figure}

It is convenient to represent graphically the fermionic
$\check{R}_{j,j+1}$-operators as in figure \ref{fig:3}.
%

Then, the full monodromy matrix $T(\lambda,u)=T_0(\lambda,u)T_1(\lambda,u)$,
defined by use of formulas (\ref{T41}), can be pictured as in
figure \ref{fig:4}, which also demonstrates the order
of writing of the operators in the product.
\def\commRb{{$\!\!\!\!\!\!{R}$}}
\def\comRbTud{{$\!\!\!\!\!\!{R}^{\iota_1\iota_2}$}}
\def\comRTd{{$\!\!\!\!\!\!\bar{R}^{\iota_2}$}}
\def\comRTu{{$\!\!\!\!\!\!\bar{R}^{\iota_1}$}}
\def\comTa{{$T_1$}}
\def\comTb{{$T_0$}}
\begin{figure}[htbp]
  \begin{center}
    \leavevmode
    
\setlength{\unitlength}{0.00033333in}
\begingroup\makeatletter\ifx\SetFigFont\undefined
\def\x#1#2#3#4#5#6#7\relax{\def\x{#1#2#3#4#5#6}}%
\expandafter\x\fmtname xxxxxx\relax \def\y{splain}%
\ifx\x\y   
\gdef\SetFigFont#1#2#3{%
  \ifnum #1<17\tiny\else \ifnum #1<20\small\else
  \ifnum #1<24\normalsize\else \ifnum #1<29\large\else
  \ifnum #1<34\Large\else \ifnum #1<41\LARGE\else
     \huge\fi\fi\fi\fi\fi\fi
  \csname #3\endcsname}%
\else
\gdef\SetFigFont#1#2#3{\begingroup
  \count@#1\relax \ifnum 25<\count@\count@25\fi
  \def\x{\endgroup\@setsize\SetFigFont{#2pt}}%
  \expandafter\x
    \csname \romannumeral\the\count@ pt\expandafter\endcsname
    \csname @\romannumeral\the\count@ pt\endcsname
  \csname #3\endcsname}%
\fi
\fi\endgroup
{\renewcommand{\dashlinestretch}{30}
\begin{picture}(17937,8550)(0,-10)
\drawline(5925,6627)(7125,7827)(7725,7227)
\drawline(4725,5427)(5325,4827)
\drawline(5218.934,4890.640)(5325.000,4827.000)(5261.360,4933.066)
\drawline(5325,4827)(5925,4227)(7125,5427)
\drawline(8325,6627)(7725,7227)
\drawline(7831.066,7163.360)(7725.000,7227.000)(7788.640,7120.934)
\drawline(8325,6627)(7125,5427)
\drawline(7188.640,5533.066)(7125.000,5427.000)(7231.066,5490.640)
\drawline(11925,5427)(13125,6627)
\drawline(13061.360,6520.934)(13125.000,6627.000)(13018.934,6563.360)
\drawline(13125,6627)(14325,7827)(14925,7227)
\drawline(11925,5427)(12525,4827)
\drawline(12418.934,4890.640)(12525.000,4827.000)(12461.360,4933.066)
\drawline(12525,4827)(13125,4227)(14325,5427)
\drawline(15525,6627)(14925,7227)
\drawline(15031.066,7163.360)(14925.000,7227.000)(14988.640,7120.934)
\drawline(15525,6627)(14325,5427)
\drawline(14388.640,5533.066)(14325.000,5427.000)(14431.066,5490.640)
\drawline(3525,4227)(2325,5427)
\drawline(2431.066,5363.360)(2325.000,5427.000)(2388.640,5320.934)
\drawline(2325,7827)(1725,7227)
\drawline(1788.640,7333.066)(1725.000,7227.000)(1831.066,7290.640)
\drawline(1725,7227)(1125,6627)(2325,5427)
\drawline(2325,7827)(3525,6627)
\drawline(3418.934,6690.640)(3525.000,6627.000)(3461.360,6733.066)
\drawline(3525,6627)(4725,5427)(4125,4827)
\drawline(3525,4227)(4125,4827)
\drawline(4061.360,4720.934)(4125.000,4827.000)(4018.934,4763.360)
\drawline(8325,1827)(9525,3027)
\drawline(9461.360,2920.934)(9525.000,3027.000)(9418.934,2963.360)
\drawline(9525,3027)(10725,4227)(11325,3627)
\drawline(8325,1827)(8925,1227)
\drawline(8818.934,1290.640)(8925.000,1227.000)(8861.360,1333.066)
\drawline(8925,1227)(9525,627)(10725,1827)
\drawline(11925,3027)(11325,3627)
\drawline(11431.066,3563.360)(11325.000,3627.000)(11388.640,3520.934)
\drawline(11925,3027)(10725,1827)
\drawline(10788.640,1933.066)(10725.000,1827.000)(10831.066,1890.640)
\drawline(10725,4227)(9525,5427)
\drawline(9631.066,5363.360)(9525.000,5427.000)(9588.640,5320.934)
\drawline(9525,7827)(8925,7227)
\drawline(8988.640,7333.066)(8925.000,7227.000)(9031.066,7290.640)
\drawline(8925,7227)(8325,6627)(9525,5427)
\drawline(9525,7827)(10725,6627)
\drawline(10618.934,6690.640)(10725.000,6627.000)(10661.360,6733.066)
\drawline(10725,6627)(11925,5427)(11325,4827)
\drawline(10725,4227)(11325,4827)
\drawline(11261.360,4720.934)(11325.000,4827.000)(11218.934,4763.360)
\drawline(7125,627)(5925,1827)
\drawline(6031.066,1763.360)(5925.000,1827.000)(5988.640,1720.934)
\drawline(5925,4227)(5325,3627)
\drawline(5388.640,3733.066)(5325.000,3627.000)(5431.066,3690.640)
\drawline(5325,3627)(4725,3027)(5925,1827)
\drawline(5925,4227)(7125,3027)
\drawline(7018.934,3090.640)(7125.000,3027.000)(7061.360,3133.066)
\drawline(4725,5427)(5925,6627)
\drawline(5861.360,6520.934)(5925.000,6627.000)(5818.934,6563.360)
\drawline(7125,3027)(8325,1827)(7725,1227)
\put(0,4227){\makebox(0,0)[lb]{\smash{{{\SetFigFont{12}{14.4}{rm}T = }}}}}
\drawline(7125,627)(7725,1227)
\drawline(7661.360,1120.934)(7725.000,1227.000)(7618.934,1163.360)
\drawline(14325,627)(13125,1827)
\drawline(13231.066,1763.360)(13125.000,1827.000)(13188.640,1720.934)
\drawline(13125,4227)(12525,3627)
\drawline(12588.640,3733.066)(12525.000,3627.000)(12631.066,3690.640)
\drawline(12525,3627)(11925,3027)(13125,1827)
\drawline(13125,4227)(14325,3027)
\drawline(14218.934,3090.640)(14325.000,3027.000)(14261.360,3133.066)
\drawline(14325,3027)(15525,1827)(14925,1227)
\drawline(14325,627)(14925,1227)
\drawline(14861.360,1120.934)(14925.000,1227.000)(14818.934,1163.360)
\dashline{60.000}(16125,1827)(17925,1827)
\dashline{60.000}(16125,6627)(17925,6627)
\drawline(1125,1827)(2325,3027)
\drawline(2261.360,2920.934)(2325.000,3027.000)(2218.934,2963.360)
\drawline(2325,3027)(3525,4227)(4125,3627)
\drawline(1125,1827)(1725,1227)
\drawline(1618.934,1290.640)(1725.000,1227.000)(1661.360,1333.066)
\drawline(1725,1227)(2325,627)(3525,1827)
\drawline(4725,3027)(4125,3627)
\drawline(4231.066,3563.360)(4125.000,3627.000)(4188.640,3520.934)
\drawline(4725,3027)(3525,1827)
\drawline(3588.640,1933.066)(3525.000,1827.000)(3631.066,1890.640)
\put(2925,6027){\makebox(0,0)[lb]{\smash{{{\SetFigFont{12}{14.4}{rm}\commRb}}}}}
\put(6525,6027){\makebox(0,0)[lb]{\smash{{{\SetFigFont{12}{14.4}{rm}\comRTd}}}}}
\put(10050,6027){\makebox(0,0)[lb]{\smash{{{\SetFigFont{12}{14.4}{rm}\commRb}}}}}
\put(13725,6027){\makebox(0,0)[lb]{\smash{{{\SetFigFont{12}{14.4}{rm}\comRTd}}}}}
\put(2925,2427){\makebox(0,0)[lb]{\smash{{{\SetFigFont{12}{14.4}{rm}\comRTu}}}}}
\put(6525,2427){\makebox(0,0)[lb]{\smash{{{\SetFigFont{12}{14.4}{rm}\comRbTud}}}}}
\put(10125,2427){\makebox(0,0)[lb]{\smash{{{\SetFigFont{12}{14.4}{rm}\comRTu}}}}}
\put(13725,2427){\makebox(0,0)[lb]{\smash{{{\SetFigFont{12}{14.4}{rm}\comRbTud}}}}}
\put(7125,8427){\makebox(0,0)[lb]{\smash{{{\SetFigFont{12}{14.4}{rm}1}}}}}
\put(9525,8427){\makebox(0,0)[lb]{\smash{{{\SetFigFont{12}{14.4}{rm}2}}}}}
\put(14325,8427){\makebox(0,0)[lb]{\smash{{{\SetFigFont{12}{14.4}{rm}3}}}}}
\put(9525,27){\makebox(0,0)[lb]{\smash{{{\SetFigFont{12}{14.4}{rm}4}}}}}
\put(7125,27){\makebox(0,0)[lb]{\smash{{{\SetFigFont{12}{14.4}{rm}3}}}}}
\put(2325,27){\makebox(0,0)[lb]{\smash{{{\SetFigFont{12}{14.4}{rm}2}}}}}
\put(16125,7152){\makebox(0,0)[lb]{\smash{{{\SetFigFont{12}{14.4}{rm}4}}}}}
\put(525,6627){\makebox(0,0)[lb]{\smash{{{\SetFigFont{12}{14.4}{rm}\comTa}}}}}
\put(525,1827){\makebox(0,0)[lb]{\smash{{{\SetFigFont{12}{14.4}{rm}\comTb}}}}}
\put(14325,27){\makebox(0,0)[lb]{\smash{{{\SetFigFont{12}{14.4}{rm}5}}}}}
\end{picture}
}

  \end{center}
  \caption{Monodromy matrix}
  \label{fig:4}
\end{figure}
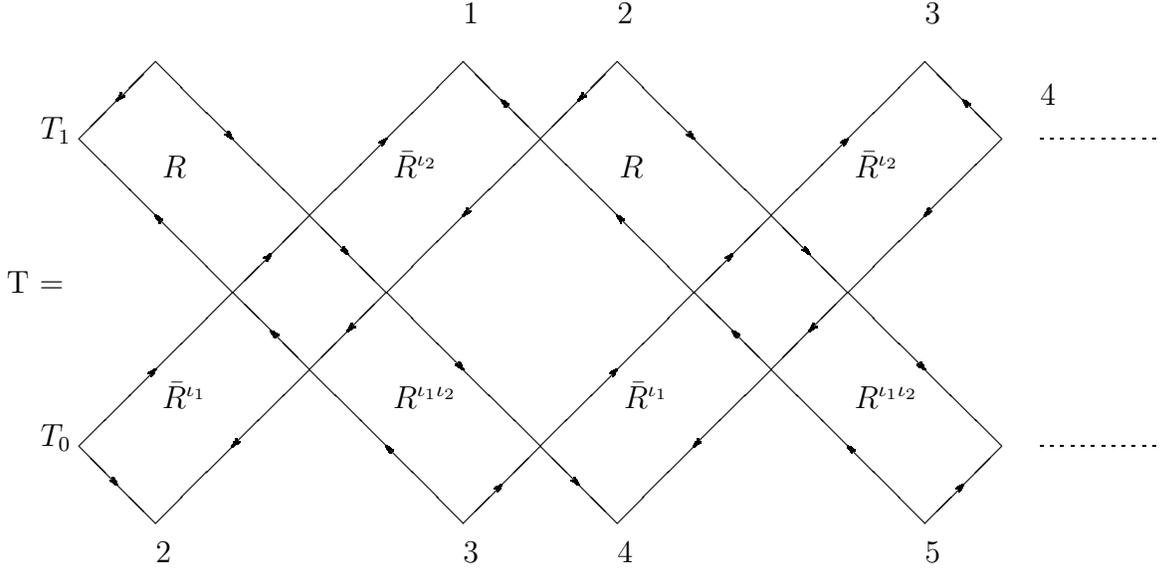


In order to be able to write a local Hamiltonian, it is necessary
to have a value 
\begin{equation}
  \label{ID}
  \check{R}_{ij}(u_0)= I_{ij}.
\end{equation}
Then the logarithmic derivative of the transfer matrix at this
point will define a local Hamiltonian
\begin{equation}
  \label{H1}
  H=-\frac{\partial \ln trT\left( u\right)}{\partial u}\mid_{u=u_0}.
\end{equation}

Technically, in order to find a Hamiltonian, one should expand 
$\check{R}_{ij}$-operators around $u_0$ and extract the first
order terms over $(u-u_0)$ in the product 
$T(\lambda,u)=T_0(\lambda,u)T_1(\lambda,u)$. In ordinary
homogeneous case by inputting the first order
expansions of the $\check{R}$-operators around unity into 
the product 
\begin{equation}
  \label{H11}
  T(\lambda,u)=\prod_{j=1}^N \check{R}_{j,j+1}(\lambda,u)
\end{equation}
it is easy to see that there is no  scattering between
long distance fermions and only the nearest-neighbour
hopping terms are contributing into the Hamiltonian.

In the present model we have a two different $R$-operators
in the product (\ref{T41}), only one of them becoming
unity for some value of the spectral parameter. 
Let us take $u_0=0$. Then,
by expanding around the point $u=0$
the $\check{R},\check{R}^{\iota_2 \iota_2}, \check{\bar{R}}^{iota_1},
\check{\bar{R}}^{\iota_2}$ operators, defined by formulas
(\ref{abcT1}),(\ref{Sol1}), (\ref{abcT2}) and (\ref{R41}), 
obtains
\begin{eqnarray}
  \label{H2}
  \check{R}_{2i,2i+1}(u)&=&
  \sin\lambda I_{2i,2i+1} +u(\cos\lambda P_{2i,2i+1}
+ c_{2i}^+ c_{2i+1} + c_{2i+1}^+ c_{2i}),\nn\\
  \check{R}^{\iota_1 \iota_2}_{2i,2i+1}(u)&=&
  \sin\lambda I_{2i,2i+1} -u(\cos\lambda P_{2i,2i+1}
+ c_i^+ c_{i+1} + c_{i+1}^+ c_{i}),\nn\\
  \check{\bar{R}}^{\iota_1}_{2i-1,2i}(u)&=&\check{R}_{2i-1,2i}(\theta)-
  u \left[\cos(\lambda+\theta) P_{2i-1,2i} -
    \cos\theta (c_{2i-1}^+ c_{2i} + c_{2i}^+ c_{2i-1})\right],\nn\\
  \check{\bar{R}}^{\iota_2}_{2i-1,2i}(u)&=&\check{R}_{2i-1,2i}(-\theta)+
  u \left[\cos(\lambda-\theta) P_{2i-1,2i} -
    \cos\eta (c_{2i-1}^+ c_{2i} + c_{2i}^+ c_{2i-1})\right],
\end{eqnarray}
where $P_{i,i+1} = n_i n_{i+1}+(1-n_i)(1-n_{i+1})$
and $\check{R}_{2i-1,2i}(\theta)$ defined by formula (\ref{R41})
with 
\begin{equation}
  \label{H3}
  a(\theta)=\sin(\lambda+\theta),\qquad b(\theta)= \sin\theta, \qquad
  c(\theta)=\sin\lambda. 
\end{equation}

It is easy to see in the Figure \ref{fig:4} that, although the 
$\check{\bar{R}}_{2i-1,2i}$-operators are not unity at $u=0$,
the Hamiltonian will still be local. The scattering due to 
$\check{\bar{R}}_{2i-1,2i}$-operators will induce an interaction 
involving at most 4 fermions. For example, the excitation
Hamiltonian $H_{23}$ for the fermions 2 and 3 (see Figure \ref{fig:4}) due to 
scatterings in $\check{\bar{R}}_{12}(\theta)$ and 
$\check{\bar{R}}_{34}(\theta)$ may create the interaction between
1, 2, 3, 4 states only.

Now, by inserting the expressions (\ref{H2}) and (\ref{H3})
for $\check{\bar{R}}^{\iota_1}, \check{\bar{R}}^{\iota_2}, 
\check{R}$ and
$\check{R}^{\iota_1 \iota_2}$
into the monodromy matrix $T(u)=T_0(u) T_1(u)$, in the linear
approximation of $u$ we obtain
\begin{equation}
  \label{TF}
  T(u) \approx (\sin^2\lambda \sin(\lambda+\theta) \sin(\lambda-\theta))^{N}
  \left(1 + u \sum_{j=1}^N H_{j}\right),
\end{equation}
with
\begin{eqnarray}
  \label{HF1}
 &&\hspace{-1cm} 
\sin\lambda \sin(\lambda+\theta)\sin(\lambda-\theta)H_{j}=\nn\\
 &&\hspace{1cm}
+(-1)^j \sin\theta \sin2\lambda (1-n_{j-1}-n_{j+2})(c^+_{j} c_{j+1}
  -c^+_{j+1} c_{j}) \nn\\
 & &\hspace{1cm}
+(-1)^{j+1}\sin^2\theta \cos\lambda\left[n_j n_{j+2}+
(1-n_j)(1-n_{j+2})\right]\nn\\
 &&\hspace{1cm}
+\sin\theta \left[\left(\sin(\lambda+(-1)^j \theta)(1-n_{j+1})+
      \sin(\lambda-(-1)^j \theta) n_{j+1}\right)c^+_{j} c_{j+2}\right.\nn\\
  &&\hspace{1cm}
- \left.\left(\sin(\lambda-(-1)^j \theta)
      (1-n_{j+1})+ \sin(\lambda +(-1)^j \theta) n_{j+1}\right)
c^+_{j+2} c_{j}\right].
\end{eqnarray}

In the limit $\lambda=\pi/2$, which corresponds to 
$XX$-model, we will have two types of 
independent  free fermions, hopping separately at the even and odd 
sites of the chain with the Hamiltonian \footnote{It seems to us that  
the model
considered in \cite{ZV} do not reproduces the free
fermionic limit at $\lambda=\pi/2$, as in case presented here.}
\begin{equation}
\label{HF}
H_{j}= \tan^2\theta\left[c_{j}^+c_{j+2} - 
c_{j+2}^+c_{j}\right].
\end{equation}

\begin{figure}[h]
  \begin{center}
    \leavevmode
    \setlength{\unitlength}{0.0015cm}
\begingroup\makeatletter\ifx\SetFigFont\undefined%
\gdef\SetFigFont#1#2#3#4#5{%
  \reset@font\fontsize{#1}{#2pt}%
  \fontfamily{#3}\fontseries{#4}\fontshape{#5}%
  \selectfont}%
\fi\endgroup%
{\renewcommand{\dashlinestretch}{30}
\begin{picture}(7954,2733)(0,-10)
\drawline(462,2304)(1362,504)(2262,2304)
        (3162,504)(4062,2304)(4962,504)
        (5862,2304)(6762,504)
\drawline(12,2304)(7212,2304)
\drawline(12,504)(7212,504)
\put(1182,54){\makebox(0,0)[lb]{\smash{{{\SetFigFont{12}{14.4}
{\rmdefault}$2j$}}}}}
\put(2982,54){\makebox(0,0)[lb]{\smash{{{\SetFigFont{12}{14.4}
{\rmdefault}$2j+2$}}}}}
\put(4782,54){\makebox(0,0)[lb]{\smash{{{\SetFigFont{12}{14.4}
{\rmdefault}$2j+4$}}}}}
\put(2082,2574){\makebox(0,0)[lb]{\smash{{{\SetFigFont{12}{14.4}
{\rmdefault}$2j+1$}}}}}
\put(3882,2574){\makebox(0,0)[lb]{\smash{{{\SetFigFont{12}{14.4}
{\rmdefault}$2j+3$}}}}}
\put(7662,2304){\makebox(0,0)[lb]{\smash{{{\SetFigFont{12}{14.4}
{\rmdefault}$s=1$}}}}}
\put(7662,504){\makebox(0,0)[lb]{\smash{{{\SetFigFont{12}{14.4}
{\rmdefault}$s=0$}}}}}
\end{picture}
} 
\end{center}
\caption{Zig-zag ladder chain}
\label{fig:5}
\end{figure}
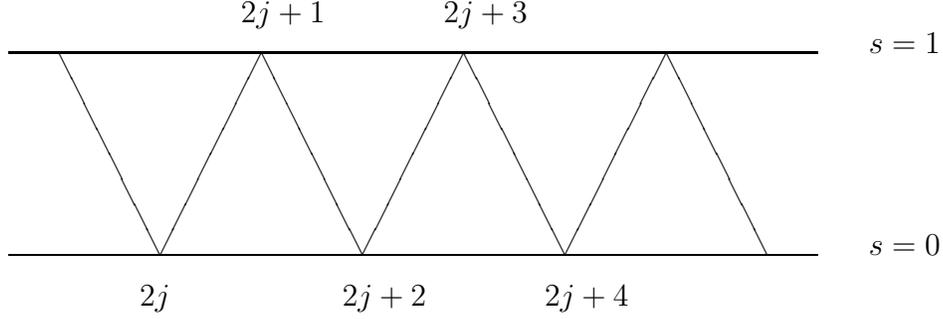

It is convenient now to write the Hamiltonian (\ref{HF1}) in a ladder form
represented graphically as in Figure \ref{fig:5}.

Let us consider the even ($2j$) and the odd ($2j+1$) points of the chain
as a sites ($j$) of two different chains labeled by $s = 0$ and $ 1$
correspondingly. The Fermi fields will be marked now as
\begin{equation}
\label{CCC}
c_{j,s}=c_{2j+s}, \qquad \qquad s = 0,1
\end{equation}

With these notations it is straightforward to obtain from the expression
(\ref{HF1}) the following ladder Hamiltonian
\begin{eqnarray}
  \label{HF2}
  &&\hspace{-1cm}
\sin(\lambda +\theta)\sin(\lambda -\theta)\sin(\lambda)H_{j,s} 
  = \nn\\
 &&\hspace{1cm} (-1)^s \sin\theta \sin2\lambda 
  \left(
    c_{j,s}^+ c_{j,s+1} - c_{j,s+1}^+ c_{j,s} 
  \right) 
  \left(
    \bar n_{j-1,s+1} - n_{j+1,s}
  \right)
  \nonumber\\
 &&\hspace{1cm}+(-1)^{s+1}\sin^2\theta\cos\lambda  
  \left[
    n_{j,s} n_{j+1,s} + (1-n_{j,s}) (1-n_{j+1,s}) 
  \right] 
  \nonumber\\
 &&\hspace{1cm}+
  \sin\theta 
  \left[
  c_{j,s}^+ c_{j+1,s}
  \left(
    \sin(\lambda + (-1)^s \theta)(1-n_{j,s+1}) +
    \sin(\lambda - (-1)^s \theta) n_{j,s+1} 
  \right)\right. 
  \nonumber\\
  &&\hspace{1cm}-c_{j+1,s}^+ c_{j,s}
  \left.\left(
    \sin(\lambda + (-1)^s \theta) n_{j,s+1} +
    \sin(\lambda - (-1)^s \theta) (1-n_{j,s+1}) 
  \right) 
  \right].
\end{eqnarray}
Hence, we have obtained a simple integrable ladder Hamiltonian,
which has a free fermionic limit.

We can also write down the Hamiltonian of the model in a spin language,
calculated by use of the ordinary representation for 
$\check{R}_{a b}^{a^{\pr}b^{\pr}}$ as
\begin{eqnarray}
\label{RR}
\check{R}_{a b}^{a^{\pr}b^{\pr}}=\left(\begin{array}{llll}
a & 0 & 0 & 0\nn\\
0 & c & b & 0\nn\\
0 & b & c & 0\nn\\
0 & 0 & 0 & a
\end{array}\right).
\end{eqnarray}
\newpage
In ladder form the result looks as follows
\begin{eqnarray}
  \label{Hspin}
 &&\hspace{-1cm}\sin(\lambda +\theta)\sin(\lambda -\theta)\sin(\lambda)H_j 
  =\nn\\
 &&\hspace{1cm} \frac{(-1)^s}{2} \sin^2 \theta \cos\lambda 
  \left[
    \sigma_{j,s}^1 \sigma_{j+1,s}^1 
    + \sigma_{j,s}^2 \sigma_{j+1,s}^2 
    - \sigma_{j,s}^3 \sigma_{j+1,s}^3 
  \right] 
  \nonumber\\
  &&+\hspace{1cm} \sin \theta \sin \lambda 
  \left\{
  \cos \lambda (-1)^s \sigma_{j,s}^3 
  \left(
    \sigma_{j,s+1}^+ \sigma_{j+1,s}^- - \sigma_{j,s+1}^- \sigma_{j+1,s}^+
  \right)\right.
  \nonumber\\
  &&-\hspace{1cm}\left.(-1)^s \cos \lambda  
  \left(\sigma_{j,s}^+ \sigma_{j,s+1}^- 
- \sigma_{j,s}^- \sigma_{j,s+1}^+\right) 
  \sigma_{j+1,s}^3 \right.\nonumber\\
  &&-\hspace{1cm}\left. \cos\theta 
  \left( 
    \sigma_{j,s}^+ \sigma_{j,s+1}^3 \sigma_{j+1,s}^- 
    - \sigma_{j,s}^- \sigma_{j,s+1}^3 \sigma_{j+1,s}^- 
  \right)
  \right\}
\end{eqnarray}

As one can see from this expression, the first term gives $XXZ$
models for the each leg of the ladder with anisotropy parameters
$\Delta=-1$ and therefore can be considered as $SU(1,1)$ 
Heisenberg chains. The other terms, which represents the
interaction between chains, are written for the each 
triangle of the ladder (see Figure \ref{fig:5}) and has a topological
form
\begin{equation}
\label{ES}
\hat{\epsilon}^{abc}\sigma_{j,s}^a \sigma_{j,s+1}^b \sigma_{j+1,s}^c,
\end{equation}
where $\hat{\epsilon}^{abc}$ is an anisotropic antisymmetric
tensor, defined by the coefficients in the formula (\ref{Hspin}).

\section{Algebraic Bethe Ansatz ($ABA$) solution of the model}
\indent

The technique of $ABA$ , called also Quantum Inverse Scattering
Method (QISM), essentially was developed in the works of Baxter
\cite{Bax} and Faddeevs group \cite{Kor}.

In order to carry out $ABA$ it is convenient to work in conventional
(not braid) formalism and use the $YBE$ for $R$-operators
in the form of formulas (\ref{R6}, \ref{R7}). Let us first define
the $L$-matrix as
\begin{eqnarray}
\label{LL}
\left(L_j(u)\right)_{a^{\pr}}^a&=&\langle a \mid R_{ij}\mid a^{\pr}
\rangle_i = (-1)^{p(a^{\pr})p(b^{\pr})}
\left(R_{ij}(u)\right)_{a^{\pr}b^{\pr}}^{a b} X_{j b}^{b^{\pr}}\nn\\
&=& \left(\begin{array}{ll}
a(u)(1-n_j)+b(u)n_j & c(u) c_j^{+}\nn\\
c(u) c_j & - a(u) n_j +b(u)(1-n_j)
\end{array}
\right),
\end{eqnarray}
which is a matrix in the horizontal auxiliary space with
operator value entities and acting on quantum space $V_j$.
The matrix elements between auxiliary states $\mid a^{\pr}\rangle$ and
 $\langle a\mid$ of the monodromy operators $T_{0,1}(\lambda,u)$
defined in (\ref{T4})
\begin{eqnarray}
\label{ABCD}
\left(T_s (u)\right)_a^{a^{\pr}}=\langle a^{\pr}\mid T_s (u)\mid a\rangle
=\left(\begin{array}{ll}
A_s(u) & B_s (u)\nn\\
C_s (u)& D_s (u)
\end{array}
\right), \qquad s=0,1
\end{eqnarray}
can be expressed as a product of the $L$-matrices as follows
\begin{eqnarray}
\label{MM}
\left(T_1\right)_{a_N}^{a_0}&=&\left(L_1\right)_{a_1}^{a_0}(u)
\left(L_2^{\iota_2}\right)_{a_2}^{a_1}(\bar u). . . . . 
\left(L_1^{\iota_2}\right)_{a_{2N}}^{a_{2N-1}}(\bar u),\nn\\
\left(T_0\right)_{a_N}^{a_0}&=&\left(L_1^{\iota_1}\right)_{a_1}^{a_0}(\bar u)
\left(L_2^{\iota_1 \iota_2}\right)_{a_2}^{a_1}(u). . . . . 
\left(L_1^{\iota_1 \iota_2}\right)_{a_{2N}}^{a_{2N-1}}(u).
\end{eqnarray}

By use of equations (\ref{R6}, \ref{R7}) one can obtain the graded $YBE$
for the monodromy matrices $T_s, \qquad s=0,1$ as follows
\begin{eqnarray}
  \label{RMM}
&&\hspace{-1cm}(-1)^{(P(b')+P(b''))P(a'')}
  {R}_{a'b'}^{ab}(u,v) 
  \left(T_0 \right)_{a''}^{a'}(u)
  \left( T_1 \right)_{b''}^{b'}(v)\nn\\
&&\hspace{1cm}=(-1)^{P(b')(P(a')+P(a))}
  \left( T_0 \right)_{b'}^{b}(v)
  \left( T_1 \right)_{a'}^{a}(u)
  \widetilde{R}_{a''b''}^{a'b'}(u,v),\nn\\
\nn\\ 
&&\hspace{-1cm}(-1)^{(P(b')+P(b''))P(a'')}
  \widetilde{R}_{a'b'}^{ab}(u,v) 
  \left( T_1 \right)_{a''}^{a'}(u)
  \left( T_0 \right)_{b''}^{b'}(v)\nn\\
&&\hspace{1cm}= (-1)^{P(b')(P(a')+P(a))}
  \left(T_1 \right)_{b'}^{b}(v)
  \left(T_0 \right)_{a'}^{a}(u)
  {R}_{a''b''}^{a'b'}(u,v). 
\end{eqnarray}

Now, following the procedure of $ABA$, let us define empty fermionic
state
\begin{equation}
\label{OO}
\mid \Omega\rangle=\prod_{i=1}^{2N}\mid 0\rangle_i
\end{equation}
as a test vacuum of the model and demonstrate that it is the eigenstate
of the transfer matrix $\tau(\lambda,u)=\tau_1(\lambda,u)\tau_0(\lambda,u)
=strT(\lambda,u)$.

The expression (\ref{LL}) for $L$-matrix shows that his action on
$\mid 0\rangle_i$ produces upper triangular matrix, therefore the action
of the monodromy matrix $\left(T_s\right)_a^{a^{\pr}}$ on 
$\mid \Omega\rangle$, defined by the formulas (\ref{ABCD}) and (\ref{MM}),
will also have upper triangular form and can be calculated easily as
\begin{eqnarray}
\label{TO}
\left(T_1(u)\right)\mid\Omega\rangle&=&\left(
\begin{array}{ll}
\left[a(u)a^{\iota_2}(\bar u)\right]^N & B_1(u)\nn\\
0 & \left[b(u)b^{\iota_2}(\bar u)\right]^N\nn\\
\end{array}\right)\mid\Omega\rangle,
\nn\\
\left(T_0(u)\right)\mid\Omega\rangle&=&\left(
\begin{array}{ll}
\left[a^{\iota_1}(\bar u)a^{\iota_1 \iota_2}(u)\right]^N & B_0(u)\nn\\
0 & \left[b^{\iota_1}(\bar u)b^{\iota_1 \iota_2}(u)\right]^N\nn\\
\end{array}\right)\mid\Omega\rangle.
\end{eqnarray}

Now it is obvious, that $\mid \Omega\rangle$ is 
eigenstate of $\tau(\lambda,u)$ with eigenvalue
\begin{eqnarray}
\label{Nu1}
\nu(u)&=&\nu_1(u)\nu_0(u),\nn\\
 \nu_0(u)&=&[a^{\iota_1}(\bar u) a^{\iota_1 \iota_2}(u)]^{N} - 
  [b^{\iota_1}(\bar u) b^{\iota_1 \iota_2}(u)]^{N},\nn\\
 \nu_1(u)&=&[a(u) a^{\iota_2}(\bar u)]^{N} - 
  [b(u) b^{\iota_2}(\bar u)]^{N}.
\end{eqnarray}
One can see from the expressions (\ref{TO}) that the operators
$C_s(u)$ act on $\mid\Omega\rangle$ as the annihilation
operators, while $B_s(u)$ act as the creation operators. That 
is why it is meaningful to look for states
\begin{equation}
\label{O2}
\mid v_1,v_2,...v_n\rangle_0=B_1(v_1)B_0(v_2)...B_x(v_n)\mid\Omega
\rangle_x,\qquad \qquad x=n(mod 2),
\end{equation}
as an $n$-particle eigenstates of $\tau(u)$ with spectral parameters
$v_i, \ i=1,...n$. In order to check whether this is true we
do not need to have an exact form of operators $B_s(u)$, we need only to
know the algebra of operators $A_s(u), D_s(u)$ and $B_s(u)$, which
can be found from the $YBE$ (\ref{RMM}) as follows
\begin{eqnarray}
  \label{AB}
  A_0(u) B_1(v) &=& - \frac{a(v-u)}{b(v-u)} B_0(v) A_1(u)
  + \frac{c(v-u)}{b(v-u)} B_0(u) A_1(v)
  \\
  A_1(u) B_0(v) &=& + \frac{a(v-u)}{b(v-u)} B_1(v) A_0(u)
  - \frac{c(v-u)}{b(v-u)} B_1(u) A_0(v)
\end{eqnarray}
and
\begin{eqnarray}
  \label{DB}
  D_0(u) B_1(v) &=& + \frac{a(u-v)}{b(u-v)} B_0(v) D_1(u)
  + \frac{c(u-v)}{b(u-v)} B_0(u) D_1(v)
  \\
  D_1(u) B_0(v) &=& - \frac{a(u-v)}{b(u-v)} B_1(v) D_0(u)
  - \frac{c(u-v)}{b(u-v)} B_1(u) D_0(v).
\end{eqnarray} 
The first terms in the right hand side of equations
(\ref{AB}) and (\ref{DB}) are so called ``wanted'' terms
and they are producing the eigenvalues $\nu(u,v_1,v_2,..v_n)$
of the state $\mid v_1,v_2,...v_n\rangle_0$ as 
\begin{eqnarray}
\label{Nu2}
&&\hspace{-1cm}
\nu(u,v_1,v_2,..v_n) = \nu_1(u,v_1,v_2,..v_n) \nu_0(u,v_1,v_2,..v_n),\nn\\
&&\hspace{-1cm}\nu_0(u,v_1,\cdots,v_{n})= 
    \frac{(-1)^{n \over 2}}{\prod_{i=1}^{n} b(v_i-u)} \cdot
    \nonumber\\
    &&\hspace{1cm} \cdot
    \left\{
      \prod_{i=1}^{n} a(v_i-u) 
      [a^{\iota_1}(\bar u) a^{\iota_1 \iota_2}(u)]^{N} - 
      \prod_{i=1}^{n} a(u-v_i) 
      [b^{\iota_1}(\bar u) b^{\iota_1 \iota_2}(u)]^{N}  
    \right\},\nn\\
    &&\hspace{-1cm}\nu_1(u,v_1,\cdots,v_{n}) = 
    \frac{(-1)^{n \over 2}}{\prod_{i=1}^{n} b(v_i-\bar u)} \cdot
    \nonumber\\
    &&\hspace{1cm} \cdot
    \left\{
      \prod_{i=1}^{n} a(v_i-\bar u) 
      [a(u) a^{\iota_2}(\bar u)]^{N} - 
      \prod_{i=1}^{n} a(\bar u-v_i) 
      [b(u) b^{\iota_2}(\bar u)]^{N}  
    \right\}.
\end{eqnarray}
But in order the state $\mid v_1,v_2,...v_n\rangle$ to be an eigenstate of 
$\tau(u)$ we need the cancellation of so called ``unwanted''
terms, produced by the second terms in the right hand side
of equations (\ref{AB}) and (\ref{DB}). This gives us the restrictions
on the spectral parameters $v_1, v_2,...v_n$ in a form of Bethe Equations
($BE$)
\begin{equation}
  \label{BE}
  \left[
    \frac{a(v_j) a^{\iota_2}(\bar v_j)}
    {b(v_j) b^{\iota_2}(\bar v_j)}
  \right]^{N}
  = - \prod_{i \neq j}^{n} \frac{a(v_j-v_i)}{a(v_i-v_j)},\qquad j=1,...n.
\end{equation}

These equations are similar, but differ by some sign factors
from the Bethe equations obtained for the model defined in \cite{ZV}
caused by the alternating anisotropy model parameter $\pm \Delta$
along the chain.

The calculation of the eigenvalues of the Hamiltonian (\ref{HF2})
is straightforward by taking of logarithmic derivatives of (\ref{Nu2})
at $u=0$ , which gives us
\begin{eqnarray}
\label{EE}
E&=& \sin\lambda \sum_{i=1}^{n}\left(\frac{1}{\sin v_i \sin(\lambda+v_i)}
-\frac{1}{\sin(v_i-\theta) \sin(\lambda+v_i-\theta)}\right)\nn\\
&+& N \frac{\sin 2\theta}{\sin(\lambda+\theta) \sin(\lambda-\theta)}.
\end{eqnarray}


\section{Open chain Hamiltonian and boundary terms of the 
model in free fermionic case ($\lambda = \pi/2$)}
\indent

In this section we follow the approach and the equations
of \cite{Ch,Sk,M,AM} to  construct the reflection $\cal K$-matrix for
our model in free fermionic case and calculate integrable 
boundary terms for the Hamiltonian.

In ordinary integrable models one defines the double row transfer 
matrix as
\begin{eqnarray}
\label{B1}
  \tau(u) &=& \mbox{tr}_0\left( K_0^+(u) T(u) 
  K_0^-(u) T(-u)^{-1}\right)\nn \\
  &=& \zeta(u)^{-L}\,\mbox{tr}_0 \left(K_0^+(u)
  \vR_{L0}(u)\vR_{L-1,L}(u)\cdots\vR_{23}(u)\vR_{12}(u) 
  \right. \nonumber\\  && \ \ \ 
  \left. \times \ K_1^-(u) 
  \vR_{12}(u)\vR_{23}(u)\cdots\vR_{L-1,L}(u)\vR_{L0}(u)\right) \; ,
\end{eqnarray}
where $N$ is the number of chain sites and $\zeta(u)$ is defined by 
the unitarity condition
\begin{equation}
\label{B2}
  \vR(u)\vR(-u) = \zeta(u)=\cos^2(u) \;
\end{equation}
for the $XX$-model.

The reflection matrices $K(u)$ (for the right side) and $K^+(u)$
(for the left side) have to obey the reflection equations, the analogues
of YBE on the boundaries, in order to ensure the commutativity
of the double row transfer matrix (\ref{B1}) for different
spectral parameters. This, together with the YBE equation, is a
sufficient  condition for integrability. The reflection equations read
\begin{equation}
\label{RK1}
  \vR_{12}(u-v) K_2^-(u) \vR_{12}(u+v) K_2^-(v) 
  =
  K_2^-(v)  \vR_{12}(u+v) K_2^-(u) \vR_{21}(u-v) 
\end{equation}
and
\begin{eqnarray}
\label{RK2}
&&  
 \vR_{12}(-u+v) K_1^+(u) 
  \vR_{12}(-u-v+\pi) K_1^+(v) =\nn\\
&& \qquad \qquad 
 = K_1^+(v)  \vR_{12}(-u-v+\pi) K_1^+(u) \vR_{12}(-u+v) \;.
\end{eqnarray}

After the fermionization (\ref{R2}),(\ref{X1}),(\ref{X2}), the solutions
of the equations (\ref{RK1}) and (\ref{RK2}) have the following form
\begin{equation}
\label{K1}
K(u)=\sin(\xi +u) n + \sin(\xi -u) (1-n)
\end{equation}
and 
\begin{equation}
\label{K2}
K^{+}(u)=-\sin(\xi^+ -u) n + \sin(\xi^+ +u) (1-n),
\end{equation}
where $\xi$ and $\xi^+$ are arbitrary parameters.

In the present model, since we already have a two-row transfer 
matrix in the closed chain case, the open chain transfer matrix 
contains four rows, with two- normal and two- backward directions.

Let us define
\begin{eqnarray}
\label{BT1}
\tau(u)=\tau_0(u) \tau_1(u)&=& \zeta(u)^{-N}tr_0 
\left[K_0^+(\theta-u)
T_0(\theta-u)K_0(\theta-u)T_0)^{-1}(-\theta+u)
\right]\times\nn\\
&\times& tr_1 \left[K_1^+(u)
T_1^{\iota_1}(u)K_1(u)T^{\iota_1}_1)^{-1}(-u)\right],
\end{eqnarray}
where $K_0(u)$, $K_1(u)$, $K^+_0(u)$ and $K^+_1(u)$ are defined by equations 
(\ref{K1}),(\ref{K2}), while $T_0(u)$ and $T_1(u)$ are defined
by (\ref{T41}) after the shift of the spectral parameter. 

Let us emphasize here, that because the boundaries of the $T_0(u)$ and
$T_1(u)$ in (\ref{T41}) can be defined in a various ways, namely,
the beginning and the end of the chain can be translated by one 
$R$-matrix, we will have a variety of boundary transfer matrices.
We now consider only one possible solution. 

The boundary transfer matrix $\tau(u)$ can be expressed graphically as in 
Figure \ref{fig:6}.
\def\coma{{$\tau(u)\ =\ $}}
\def\comb{{$K^+_1(u)$}}
\def\comc{{$K^+_0(\theta-u)$}}
\def\comd{{$\left(T_1^{\iota_1}\right)^{-1}(-u)$}}
\def\come{{$T_1^{\iota_1}(u)$}}
\def\comf{{$T_0^{-1}(-\theta+u)$}}
\def\comg{{$T_0(\theta-u)$}}
\def\comh{{$K_1(u)$}}
\def\comi{{$K_0(\theta-u)$}}
\begin{figure}[htbp]
    \leavevmode

\setlength{\unitlength}{0.0004in}
\begingroup\makeatletter\ifx\SetFigFont\undefined
\def\x#1#2#3#4#5#6#7\relax{\def\x{#1#2#3#4#5#6}}%
\expandafter\x\fmtname xxxxxx\relax \def\y{splain}%
\ifx\x\y   
\gdef\SetFigFont#1#2#3{%
  \ifnum #1<17\tiny\else \ifnum #1<20\small\else
  \ifnum #1<24\normalsize\else \ifnum #1<29\large\else
  \ifnum #1<34\Large\else \ifnum #1<41\LARGE\else
     \huge\fi\fi\fi\fi\fi\fi
  \csname #3\endcsname}%
\else
\gdef\SetFigFont#1#2#3{\begingroup
  \count@#1\relax \ifnum 25<\count@\count@25\fi
  \def\x{\endgroup\@setsize\SetFigFont{#2pt}}%
  \expandafter\x
    \csname \romannumeral\the\count@ pt\expandafter\endcsname
    \csname @\romannumeral\the\count@ pt\endcsname
  \csname #3\endcsname}%
\fi
\fi\endgroup
{\renewcommand{\dashlinestretch}{30}
\begin{picture}(12054,5550)(0,-10)
\drawline(7605,5067)(4905,5067)(4005,4167)
        (4905,3267)(7605,3267)
\drawline(7485.000,3237.000)(7605.000,3267.000)(7485.000,3297.000)
\drawline(7605,522)(10305,522)(11205,1422)
        (10305,2322)(7605,2322)
\drawline(7725.000,2352.000)(7605.000,2322.000)(7725.000,2292.000)
\drawline(7605,2322)(4905,2322)(4005,1422)
        (4905,522)(7605,522)
\drawline(7485.000,492.000)(7605.000,522.000)(7485.000,552.000)
\put(6525,5427){\makebox(0,0)[lb]{\smash{{{\SetFigFont{12}{14.4}{rm}\comd}}}}}
\put(6525,3627){\makebox(0,0)[lb]{\smash{{{\SetFigFont{12}{14.4}{rm}\come}}}}}
\drawline(7605,3267)(10305,3267)(11205,4167)
        (10305,5067)(7605,5067)
\drawline(7725.000,5097.000)(7605.000,5067.000)(7725.000,5037.000)
\put(6525,1827){\makebox(0,0)[lb]{\smash{{{\SetFigFont{12}{14.4}{rm}\comf}}}}}
\put(11700,1467){\makebox(0,0)[lb]{\smash{{{\SetFigFont{12}{14.4}{rm}\comi}}}}}
\put(6570,27){\makebox(0,0)[lb]{\smash{{{\SetFigFont{12}{14.4}{rm}\comg}}}}}
\put(1845,1422){\makebox(0,0)[lb]{\smash{{{\SetFigFont{12}{14.4}{rm}\comc}}}}}
\put(1845,4122){\makebox(0,0)[lb]{\smash{{{\SetFigFont{12}{14.4}{rm}\comb}}}}}
\put(0,2772){\makebox(0,0)[lb]{\smash{{{\SetFigFont{12}{14.4}{rm}\coma}}}}}
\put(11655,4167){\makebox(0,0)[lb]{\smash{{{\SetFigFont{12}{14.4}{rm}\comh}}}}}
\end{picture}
}
  \caption{Boundary transfer matrix}
  \label{fig:6}
\end{figure}
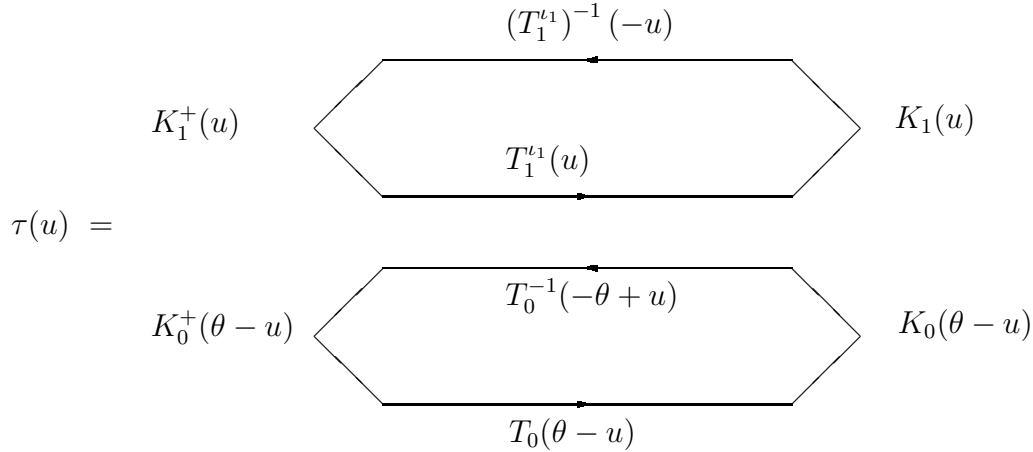

This way of writing the boundary transfer matrix $\tau(u)$ is 
not convenient for the calculation of the Hamiltonian, particularly
for the bulk part, but it is convenient to analyze its commutation
properties  for different spectral parameters. It is clear
that the boundary YBEs (reflection equations) are the same as
(\ref{RK1}) and (\ref{RK2}) for the ordinary case. The expressions
(\ref{K1}) and (\ref{K2}), after the shift of the spectral parameter
$u \rightarrow \theta-u$ for the
0 row  will ensure the commutativity
of the boundary transfer matrices (\ref{BT1}). Moreover, by use
of YBE (\ref{R6}) and (\ref{R7}) we can translate the 
$T_0^{-1}(-\theta+u)$ row in  Figure \ref{fig:6} to very top
by commuting with $T_1^{\iota_1}(u)$ and 
$(T_1^{\iota_1})^{-1}(-u)$ rows and bring it to the  
form, convenient for the calculation of the Hamiltonian:
\begin{equation}
\label{BT2}
\tau(u)=tr\left[{\cK}^+(u) T_0(\theta-u)T_1(u)\cK(u)
T^{-1}_1(-u)T^{-1}_0(-\theta+u)\right]\;.
\end{equation}
Here the effective reflection matrices ${\cK}^+(u), \cK(u)$
have been introduced and are defined as follows
\begin{eqnarray}
\label{K3}
\cK(u)&=&\vR (2u +\theta) K_1(u)\vR (\theta)
K_0(\theta -u),\nonumber\\
{\cK}^{+}(u)&=&-\vR(\theta-2u)K_1^{+}(u)\vR (-\theta) K_0^+(\theta- u).
\end{eqnarray}

It is clear, that the boundary transfer matrix $\tau(u)$ 
can be represented graphically as in Figure \ref{fig:7}.
\def\coma{{$\scriptstyle \tau(u)\ =\ $}}
\def\comb{{$\scriptstyle K^+_1(u)$}}
\def\comc{{$\scriptstyle K^+_0(\theta-u)$}}
\def\comd{{$\scriptstyle \left(T_1^{\iota_1}\right)^{-1}(-u)$}}
\def\come{{$\scriptstyle T_1^{\iota_1}(u)$}}
\def\comf{{$\scriptstyle T_0^{-1}(-\theta+u)$}}
\def\comg{{$\scriptstyle T_0(\theta-u)$}}
\def\comh{{$\scriptstyle K_1(u)$}}
\def\comi{{$\scriptstyle K_0(\theta-u)$}}
\begin{figure}[htbp]
  
\setlength{\unitlength}{0.00025in}
\begingroup\makeatletter\ifx\SetFigFont\undefined
\def\x#1#2#3#4#5#6#7\relax{\def\x{#1#2#3#4#5#6}}%
\expandafter\x\fmtname xxxxxx\relax \def\y{splain}%
\ifx\x\y   
\gdef\SetFigFont#1#2#3{%
  \ifnum #1<17\tiny\else \ifnum #1<20\small\else
  \ifnum #1<24\normalsize\else \ifnum #1<29\large\else
  \ifnum #1<34\Large\else \ifnum #1<41\LARGE\else
     \huge\fi\fi\fi\fi\fi\fi
  \csname #3\endcsname}%
\else
\gdef\SetFigFont#1#2#3{\begingroup
  \count@#1\relax \ifnum 25<\count@\count@25\fi
  \def\x{\endgroup\@setsize\SetFigFont{#2pt}}%
  \expandafter\x
    \csname \romannumeral\the\count@ pt\expandafter\endcsname
    \csname @\romannumeral\the\count@ pt\endcsname
  \csname #3\endcsname}%
\fi
\fi\endgroup
{\renewcommand{\dashlinestretch}{30}
\begin{picture}(11694,7530)(0,-10)
\drawline(7245,6057)(4545,6057)(3645,4257)
        (4995,1557)(7245,1557)
\drawline(7125.000,1527.000)(7245.000,1557.000)(7125.000,1587.000)
\drawline(7245,612)(9945,612)(10845,2457)
        (8595,6957)(7245,6957)
\drawline(7365.000,6987.000)(7245.000,6957.000)(7365.000,6927.000)
\drawline(7245,6957)(5895,6957)(3645,2457)
        (4545,612)(7245,612)
\drawline(7125.000,582.000)(7245.000,612.000)(7125.000,642.000)
\put(6345,2007){\makebox(0,0)[lb]{\smash{{{\SetFigFont{12}{14.4}{rm}\come}}}}}
\put(6255,27){\makebox(0,0)[lb]{\smash{{{\SetFigFont{12}{14.4}{rm}\comg}}}}}
\drawline(7245,1557)(9495,1557)(10845,4257)
        (9945,6057)(7245,6057)
\drawline(7365.000,6087.000)(7245.000,6057.000)(7365.000,6027.000)
\put(11295,4257){\makebox(0,0)[lb]{\smash{{{\SetFigFont{12}{14.4}{rm}\comh}}}}}
\put(5445,5157){\makebox(0,0)[lb]{\smash{{{\SetFigFont{12}{14.4}{rm}\comd}}}}}
\put(11295,2457){\makebox(0,0)[lb]{\smash{{{\SetFigFont{12}{14.4}{rm}\comi}}}}}
\put(1395,4257){\makebox(0,0)[lb]{\smash{{{\SetFigFont{12}{14.4}{rm}\comb}}}}}
\put(1350,2457){\makebox(0,0)[lb]{\smash{{{\SetFigFont{12}{14.4}{rm}\comc}}}}}
\put(0,3357){\makebox(0,0)[lb]{\smash{{{\SetFigFont{12}{14.4}{rm}\coma}}}}}
\put(6345,7407){\makebox(0,0)[lb]{\smash{{{\SetFigFont{12}{14.4}{rm}\comf}}}}}
\end{picture}
}
\hspace{.5cm}
\begingroup\makeatletter\ifx\SetFigFont\undefined
\def\x#1#2#3#4#5#6#7\relax{\def\x{#1#2#3#4#5#6}}%
\expandafter\x\fmtname xxxxxx\relax \def\y{splain}%
\ifx\x\y   
\gdef\SetFigFont#1#2#3{%
  \ifnum #1<17\tiny\else \ifnum #1<20\small\else
  \ifnum #1<24\normalsize\else \ifnum #1<29\large\else
  \ifnum #1<34\Large\else \ifnum #1<41\LARGE\else
     \huge\fi\fi\fi\fi\fi\fi
  \csname #3\endcsname}%
\else
\gdef\SetFigFont#1#2#3{\begingroup
  \count@#1\relax \ifnum 25<\count@\count@25\fi
  \def\x{\endgroup\@setsize\SetFigFont{#2pt}}%
  \expandafter\x
    \csname \romannumeral\the\count@ pt\expandafter\endcsname
    \csname @\romannumeral\the\count@ pt\endcsname
  \csname #3\endcsname}%
\fi
\fi\endgroup
{\renewcommand{\dashlinestretch}{30}
\begin{picture}(11199,7800)(0,-10)
\drawline(6750,7227)(4050,7227)(3150,5427)
        (5400,927)(6750,927)
\drawline(6630.000,897.000)(6750.000,927.000)(6630.000,957.000)
\drawline(6750,1782)(9450,1782)(10350,3627)
        (9000,6327)(6750,6327)
\drawline(6870.000,6357.000)(6750.000,6327.000)(6870.000,6297.000)
\drawline(6750,6327)(4500,6327)(3150,3627)
        (4050,1782)(6750,1782)
\drawline(6630.000,1752.000)(6750.000,1782.000)(6630.000,1812.000)
\put(10800,5427){\makebox(0,0)[lb]{\smash{{{\SetFigFont{12}{14.4}{rm}\comi}}}}}
\put(10800,3627){\makebox(0,0)[lb]{\smash{{{\SetFigFont{12}{14.4}{rm}\comh}}}}}
\drawline(6750,927)(8100,927)(10350,5427)
        (9450,7227)(6750,7227)
\drawline(6870.000,7257.000)(6750.000,7227.000)(6870.000,7197.000)
\put(0,3627){\makebox(0,0)[lb]{\smash{{{\SetFigFont{12}{14.4}{rm}=}}}}}
\put(4950,5652){\makebox(0,0)[lb]{\smash{{{\SetFigFont{12}{14.4}{rm}\comd}}}}}
\put(1035,5427){\makebox(0,0)[lb]{\smash{{{\SetFigFont{12}{14.4}{rm}\comc}}}}}
\put(1080,3627){\makebox(0,0)[lb]{\smash{{{\SetFigFont{12}{14.4}{rm}\comb}}}}}
\put(5580,27){\makebox(0,0)[lb]{\smash{{{\SetFigFont{12}{14.4}{rm}\comg}}}}}
\put(5625,2277){\makebox(0,0)[lb]{\smash{{{\SetFigFont{12}{14.4}{rm}\come}}}}}
\put(5625,7677){\makebox(0,0)[lb]{\smash{{{\SetFigFont{12}{14.4}{rm}\comf}}}}}
\end{picture}
}

  \caption{Identity relation for boundary transfer matrix}
  \label{fig:7}
\end{figure}
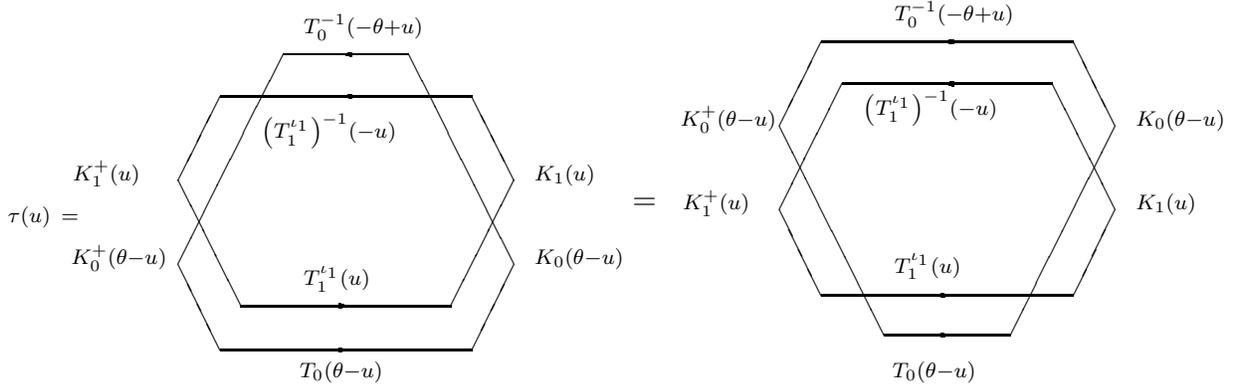

The equation in Figure \ref{fig:7} is a consequence of the reflection
equations (\ref{RK1}) and (\ref{RK2}).

Now again, in order
to calculate the boundary terms,  we should input
into the expression of the boundary transfer matrix (\ref{BT2}), 
as in the case of the bulk Hamiltonian, 
the linear expansions of the operators $\check{R} , \check{\bar{R}}$, 
defined
by the formulas (\ref{H2} - \ref{H3}) and
the following linear expansions of 
the reflection $K$-matrices around $u=0$
\begin{eqnarray}
\label{EK1}
K_i(u)&=& \sin\xi +u \cos\xi\left[n_i -(1 - n_i)\right]\; ,\nn\\
K_i(\theta-u)&=&\left[\sin(\xi+\theta)n_i +
\sin(\xi -\theta)(1-n_i)\right]+\nn\\
&+& u\left[-\cos(\xi+ \theta) n_i +\cos(\xi -\theta)(1 - n_i)\right]
\end{eqnarray}
and extract the terms linear in $v$. After some
calculations one can obtain 
\begin{equation}
\label{HB3}
H= \sum_{j=1}^{N} 2 H_{j} + H_R + H_L,
\end{equation}
where $H_{j}$ is the bulk Hamiltonian, which have been
calculated in the previous section and given by the formula
(\ref{HF}), while the right -$H_R$ and the left- $H_L$ boundary
terms have the following expressions
\begin{eqnarray}
\label{HRL}
H_R&=&\tan\theta\left(\left[{\sin(\xi-\theta) \over \sin(\xi+\theta)}+1\right]
c^+_{N}c_{N-2} -\left[{\sin(\xi+\theta) \over \sin(\xi-\theta)}+1\right]
c^+_{N-2}c_{N}\right)\nn\\
&+&{1 \over \cos\theta}\left(\left[{\sin(\xi-\theta) \over \sin(\xi+\theta)}-1
-2\tan\theta\right]
c^+_{N-1}c_{N} +\left[{\sin(\xi+\theta) \over \sin(\xi-\theta)}-1
+2 \tan\theta\right] c^+_{N}c_{N-1}\right)\nn\\
&+&\left[{\cot\xi \over \cos^2\theta}-\cot(\xi+\theta)\right]n_N +
-\left[{\cot\xi \over \cos^2\theta}-\cot(\xi-\theta)\right](1-n_N)\nn\\
&-&\tan^2\theta \cot\xi n_{N-1} +\tan^2\theta \cot\xi (1-n_{N-1})-
2\tan\theta,\nn\\
H_L&=&-4{\sin\theta \over \sin\xi \cos2\theta}=const
\end{eqnarray}

It is important to mention that, besides the translation of the
boundary of the chain, there is another possibility for a modification
of the boundary transfer matrix (\ref{BT1}). We can take the return
paths $T_0^{-1}(-\theta+u)$ and $(T_1^{\iota_1})^{-1}(-u)$
not above the direct chain, as it was considered and presented in the
Figure \ref{fig:6}, but below. Then it is not hard to see that this will
interchange the left and right boundary terms (\ref{HRL}) in the
Hamiltonian. Finally, there are four integrable models with
different boundary terms, only one of which being explicitly
written here.


\section*{Acknowledgment}
\indent
The authors R.~P. and A.~S. acknowledge LAPTH for the warm hospitality
during this work.


\end{document}